# 基于高密度激光点云的道路破损检测

# Pavement Crack Detection Based on Mobile Laser Scanning Data

武汉大学测绘学院

王猛

# 摘　要

　　道路破损是影响道路通行能力的主要因素之一。目前，我国公路通车里程已位居世界第一，使用传统的人工方法进行道路病害检测费时费力，且检测结果易受检测人员的主观影响。而使用数字图像进行道路破损检测容易受到光照阴影和周围环境的影响，导致检测精度大大降低。因此，研究一种新型道路破损检测方法迫在眉睫具有很大的应用价值。

　　本文提出基于高密度激光点云进行道路破损检测。高密度激光点云可由车载激光扫描系统获得，其在车载平台中集成 GNSS/INS、激光扫描仪以及相机等传感器，在平台的高速运动中自动获取道路及道路周边地物的三维空间信息，是目前最先进的三维空间信息获取技术之一。车载激光扫描系统在获取高密度激光点云时不受光照等干扰，且获取速度快，能大大加快作业效率。

　　本文提出的道路破损检测方案分为数据准备、图像预处理、二值化分割和道路破损增强四个部分，将数字图像易于处理和激光点云数据质量高的优势结合来解决问题。高密度激光点云首先被内插为道路面特征图像，接着，中值滤波、灰度形态学、局部自适应阈值以及基于多尺度迭代的张量投票方法被用来从道路面特征图像中提取出道路破损信息。最后，本文使用了 Hausdorff 距离评定检测精度，SM 值可达 95 左右，可见道路破损被很好地检测了出来，表明本文提出的道路破损检测方法可以较好地服务于公路管理部门的道路破损检测。

**关键词**：道路破损；道路病害检测；激光点云；灰度形态学；计算机视觉；数字图像处理；局部自适应阈值；张量投票；Hausdorff 距离

# ABSTRACT

Pavement cracks is one of the most important reasons that affects the road capacity. Nowadays, China has the longest highway mileage in the world, thus using traditional manual methods to detect pavement cracks is both time and labor consuming, besides, the detection results are prone to be affected by detectors, which is often subjective. Meanwhile, using digital image to detect pavement cracks may be affected by illumination and shadows, which could dramatically reduce the detection precision. Therefore, designing a new detection method has important significance.

This paper proposes a new method of detecting pavement cracks using high density laser point cloud. High density laser point cloud can be gathered through Vehicle-borne laser scanning system, which integrates a variety types of sensors which include GNSS/INS，laser scanner and cameras. It can automatically collect 3-D spatial information around it in a high speed, it's one of the most advanced 3-D spatial information acquisition technology. The system is not affected by illumination while gathering laser point cloud, besides, it gathers laser point cloud very fast, which greatly improves the detection efficiency.

The method proposed consists of four parts, which are data preparation, image preprocessing, binarization and crack enhancement. This method combines the advantages of digital image and laser point cloud to solve the problem. High density laser point cloud are first interpolated into georeferenced feature (GRF) image, then median filter, morphology, local adaptive threshold and multi scale iterative tensor voting method are used to detect pavement cracks from GRF image. At last, Hausdorff distance is used to evaluate detection precision. The SM value reached around 95, indicates that pavement cracks are well detected and the method proposed can serve the municipal departments well to detect pavement cracks.

# 目　录

# 第一章 绪论

## 1.1 研究背景

随着大规模的城市化进程正在我国逐步推进，我国的公路总里程也在不断增长，尤其是高等级公路和城市市区道路，根据交通部发布的统计数据，截止 2016 年底，我国高速公路已超过 16 万公里，总里程位居世界首位。

道路的建设极大地促进了社会和经济发展，但是道路建设后的保养和维护成为了相当费时费力的问题，需要定期对道路路面进行检查，发现道路病害并及时修缮。道路病害若不能够及时处理，则会对行车安全造成巨大的隐患，还会缩短道路寿命。若能在道路病害发生早期及时发现，安全隐患就会及时被排查，道路的维护费用也会显著降低。所以道路养护的重要性不亚于建设新的公路。

目前，路面数据中重要指标之一路面破损主要采用两种手段来检测，分别是人工实地检测以及使用影像进行检测。[1]

人工实地检测工作量大、效率低，检测中影响正常交通，存在安全隐患。并且，人工实地检测容易受到检测员的主观因素影响，这样就不利于对道路病害进行客观准确的检测。并且，采用人工实地检测方式，会导致道路信息更新周期较长。因此用人工对路况进行检测会成为一项越来越繁重的工作。如果采用自动化的方法去提取道路病害信息，则可以极大地减少人力物力的消耗，同时提高道路破损监测的效率和准确率。使用影像进行道路破损检测极大地提高了检测的效率和自动化程度，但是影像容易受到光照和阴影的影响，这会导致识别精度不高甚至误识别，其提取精度和精度要求较为严格的生产应用仍有一定距离。

车载激光点云扫描系统作为一种先进的测量手段，集成了激光雷达系统（LiDAR）和 GNSS/INS 系统，它能够采集 LiDAR 扫描范围内物体的高密度激光点云，具有快速、不与测量物接触、实时、动态、主动、高密度及高精度等特点。[2]例如，配备有该系统的采集车能够一次采集到整个路面和道路周围物体，比如树木、房屋以及交通标志牌的高密度激光点云。同时，车载激光点云扫描系统所配备的高脉冲激光雷达使得其采集到的激光点云密度足够大，能够采集到道路

面破损，因此其在道路维护和交通领域有很大的应用潜力。

车载激光点云扫描系统采集到的高密度激光点云中包含有每个点的反射强度信息，这些强度信息已经被广泛利用在交通标志牌、道路标线和路坎的提取中。而在道路破损中，破损处的激光点云反射强度相比与道路面其它部分的反射强度小了很多，因此激光点云的反射强度信息可以被用来从道路面中提取道路破损。此外，相较于传统的光学图像，高密度激光点云的采集不会受到光照和阴影的影响，因此车载激光点云扫描系统可以 24 小时作业，采集车可以在车流量较少的夜间作业，以此来消除车流对道路破损检测造成的影响。

道路面破损信息的自动高效提取是公路管理与维护中的核心技术之一，使用车载激光扫描技术从车载激光点云中提取道路面破损信息能够有效地提高道路面破损信息提取的效率，降低道路维护管理的成本。因此，本项目研究车载激光点云数据道路破损信息检测提取技术，具有重要的实际意义。

## 1.2 道路破损概述

我国公路路面主要是沥青路面。这些年来，沥青作为优良的道路面铺设材料已被广泛使用在了许多城乡公路的铺设上。沥青路面在通车使用一段时间之后会由于行车压力和环境因素产生各种破损。

沥青路面破损分为两类：结构性破损和功能性破损。结构性破损会造成路面各层的承载能力降低，反映在表面上往往是裂缝；功能性破损会造成路面提供给道路用户的服务能力下降，反映在路面上则是平整度降低和车辙加深。[3]常见的结构性破损有龟状裂缝、块状裂缝、纵向裂缝和横向裂缝等。常见的功能性破损有不平整、坑洞、拥包、车辙等。[4]

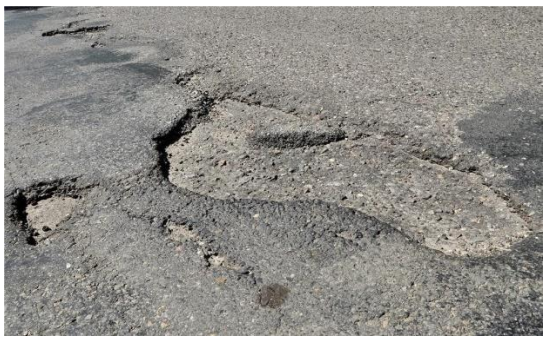
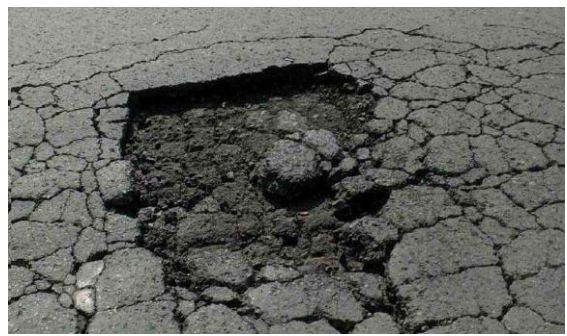

图 1.1 功能性破损　　　　　　　图 1.2 功能性破损

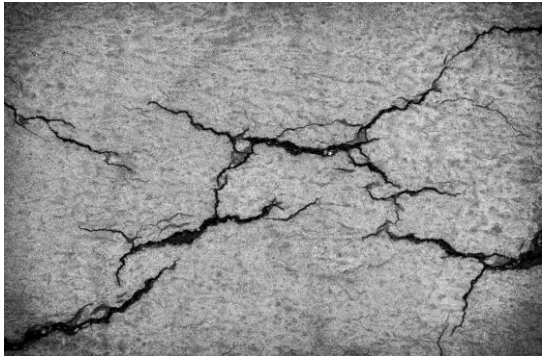 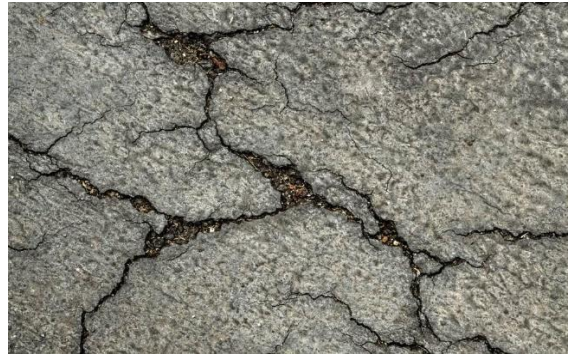

图 1.3 结构性破损　　　　　　　　　图 1.4 结构性破损

经路政部门的分析，造成道路面破损的主要原因有：超负荷的交通量与车辆超载、变暖的气候、沥青的质量、泛油和水破坏等。

## 1.3　道路破损自动检测的技术进展

随着科学技术的不断发展，道路破损的自动检测技术也在不断进步。许多学者利用不同学科不同领域的技术提出了许多自动检测提取道路破损的方法，也有不少企业开发了整套道路病害采集系统，如美国的"PCES"系统、瑞典的"PAVUE"系统、日本的"Komatsu"系统和瑞士的"CREHOS"系统等。[5]按照提取手段的数据源划分，主要分为基于图像和目前新出现的基于激光点云。

### 1.3.1　基于图像的道路破损自动检测

自从二十世纪八十年代起，摄影技术开始越来越多地应用在道路破损自动检测领域。其主要思想是使用扫描摄像机（CCD/CMOS）采集道路面的数字图像，接着利用数字图像处理技术从采集到的道路面图像中提取道路裂缝。

经过研究人员的不断研究，目前基于图像的道路破损自动检测方法主要围绕着三大部分进行：图像增强、图像分割和提取优化。图像增强是为了改善道路面图像明暗不均的情况，使其有利于图像分割。图像分割则是利用二值化，将道路破损和道路面进行有效分离，从而实现道路破损的提取。提取优化是基于自动分割的结果，并根据道路破损的特征，如几何特征，对分割结果进行优化提高的过程。自动检测的方法不同主要是研究人员在上述三大部分中使用的方法不同。

主流的方法有如下几种：李刚，贺昱曜等人提出使用大津法（OTSU）和最大互信息量结合的图像分割算法，并依据轮廓跟踪原理，得到路面破损区域的面积、

周长等定量参数。[6]Bhagvati 等人将形态学的概念引入到了道路面破损检测。[7]Siriphan Jitprasithsiri 在对图像增强时，使用了传统的中值滤波方法。[8]Velinsk 等人利用直方图分析的方法进行图像的分割，该方法对于路面破损较为明显的区域较为有效。[9]近些年随着人工智能方面的技术蓬勃发展，神经网络和机器学习技术也广泛应用于道路破损提取方面，Meignen 等人将神经网络模型引入到道路图像的处理，基于获得的样本进行训练，从而对采集到的图像中的道路病害进行判断，取得较高的准确性。[10]

但是上述方法的成功与否很大程度上取决于数据采集质量的好坏，也就是道路面图像的质量。由于光照的影响，会造成采集到的道路图像明暗不均，这会给道路破损的提取造成很大困难，甚至导致图片信息丢失。此外，由于道路图像很多是由路面检测车动态采集的，这也会导致图像模糊和图像分辨率不均一等问题，使得道路破损不能被定量检测。另外路面本身存在的污渍等也会造成道路面图像质量的下降。

### 1.3.2 基于激光点云的道路破损自动检测

几年来，移动激光雷达技术（LiDAR）蓬勃发展，其应用也越来越广泛。车载 LiDAR 目前主要用于提取路面以及道路周边信息，例如建筑物的立面信息、道路附属设施(路灯)、交通标示牌、树木、道路标线等等，而对提取道路破损的研究比较少。李爱霞和管海燕等人以车载 LiDAR 数据为数据源，通过提取道路路坎分割激光点云，提取道路面点云，并将点云数据转化为二维特征图像。基于转化后的图像，利用张量投票方法，从图像中提取道路裂缝。[11]本文将按照这种思路出发，尝试对基于高密度激光点云的道路破损自动检测技术进行改进和创新。

# 第二章 基于高密度激光点云的道路破损检测过程

本文的提取方法与李爱霞和管海燕等人提出的《基于张量投票的道路表面裂缝检测》的提取方法[11]有相似之处，即通过将道路面激光点云转换为二维特征图像，并在特征图像上检测道路破损，从而将激光雷达与图像的优点结合起来进行检测道路破损的工作。从检测流程来看，其一共分为四个步骤，分别为：数据准备、图像预处理、二值化分割和道路破损增强。

- 数据准备：数据准备是将高密度激光点云转化为道路面的二维特征图像，即使用车载激光扫描系统获取道路及道路周边点云，对原始车载激光点云进行滤波，得到道路面点云，并使用高分辨率二维特征影像插值算法将道路面点云转化得到道路面二维特征图像。

- 图像预处理：图像预处理是指对特征图像进行优化处理，从而减少图像亮度不均和道路面的干扰物对破损检测造成的干扰的过程。本文首先使用中值滤波去除一定量的噪声，接着使用灰度形态学中底帽变换的方法对道路面二维特征图像进行图像均衡，使特征图像亮度均一，同时排除道路标线对提取结果的影响，从而有利于检测道路面破损。

- 二值化分割：大部分基于图像的道路面破损检测方法使用了全局阈值进行图像二值化分割，但由于道路图像的采集十分受光照的影响，因此反映在道路图像上则为图像亮度不均。所以使用全局阈值进行二值化分割的效果并不理想。同理，由于车载激光点云的强度也会受到扫描距离、扫描入射角、物体的几何性质等因素的影响，因此无法使用全局阈值对道路面二维特征图像进行二值化，本文研究使用局部自适应阈值对道路面二维特征图像进行二值化分割，获得带有噪声的道路破损图像，从而得到初步检测结果。

- 道路破损增强：由于道路面有水渍、污渍和黑色斑点的存在，所以使用局部自适应阈值进行二值化分割得到的道路面破损图像带有噪声。为了解决这个问题，本文研究结合道路裂缝的几何信息，采用基于多尺度迭代的张量投票算法剔除二值化结果中的噪声，并对提取出的道路破损进行增强，得到最终的检测结果。

为了加深读者对本文的检测方法的理解以及更清晰地展示流程步骤，现将本文的基于高密度激光点云的道路破损检测流程展示如下图：

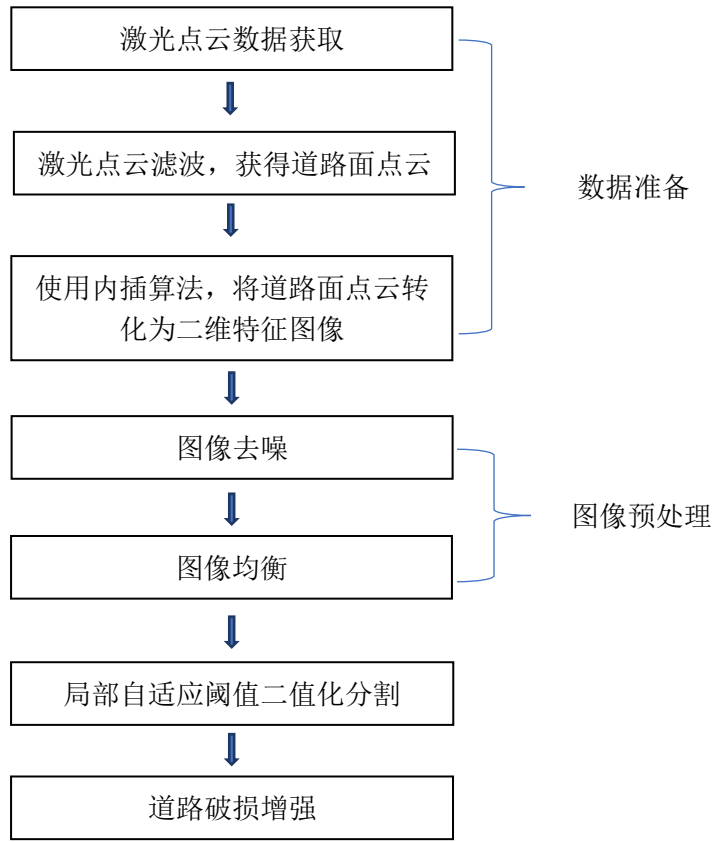

图 2.1 道路破损检测流程

# 第三章 数据准备

本文实验与测试使用的数据集是由车载移动激光雷达采集到的点云内插之后得到的高分辨率道路面特征图像。图 3.1 为装有车载 LiDAR 系统的采集车辆。该系统由 SPAN-SE LCI 组合导航定位系统、VUX-1 激光扫描仪和全景相机组成，它们被集成到采集车辆上。为了在后处理阶段融合数据，所有的传感器的时间都与 GPS 时间同步，且激光扫描仪与惯导系统的视准轴误差也在采集数据前被校准。图 3.2 为采集到的高密度激光点云数据。

在激光点云采集完成后，主要通过激光点云的高程信息和路肩的位置来提取道路面的激光点云，并用内插的方法将道路面的激光点云内插为高分辨率道路面特征图像。如何内插得到高分辨率道路面特征图像非本文重点，在此不再赘述。图 3.3 中红色部分为提取出的道路面激光点云，图 3.4 为经内插后得到的高分辨率道路面特征图像。本文采用了陈长军导师团队内插处理后的道路面特征图像作为数据集进行实验和测试。

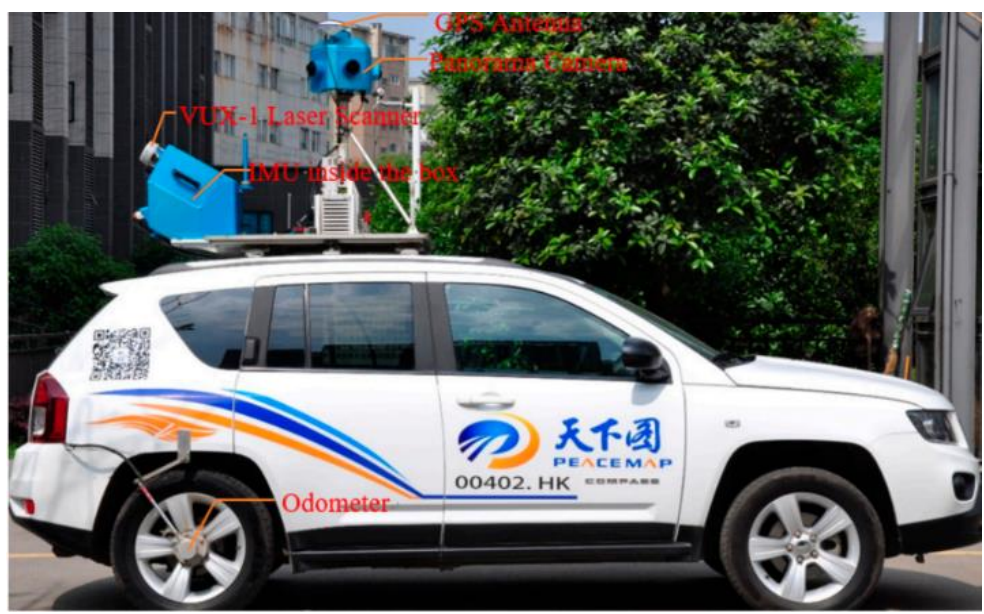

图 3.1 高密度激光点云采集车辆

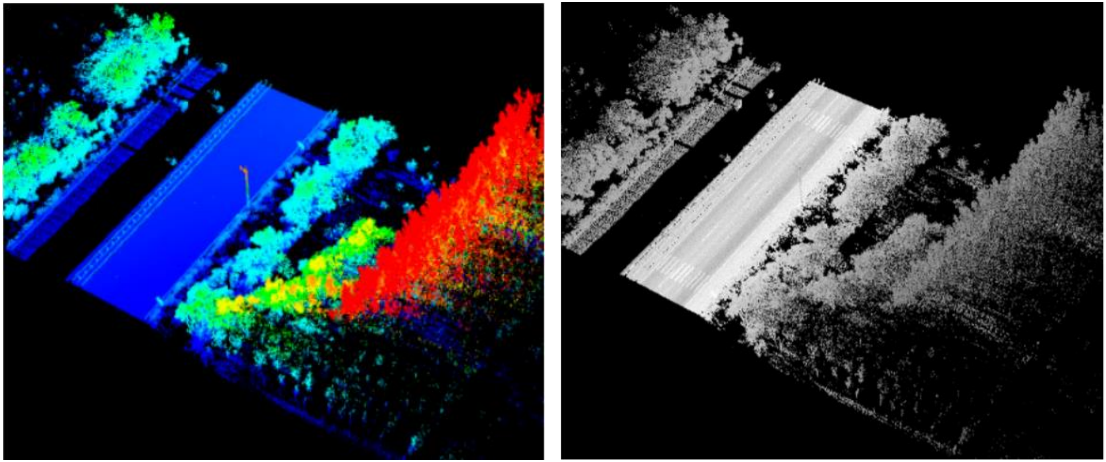

(a)根据反射点高程渲染的激光点云数据　(b)根据反射强度渲染的激光点云数据

图 3.2 高密度激光点云数据

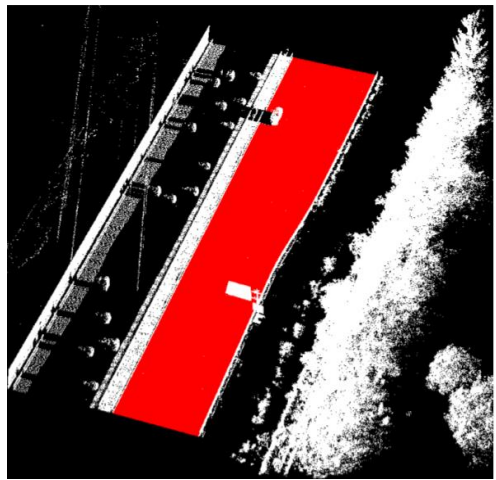

图 3.3 提取出的道路面激光点云

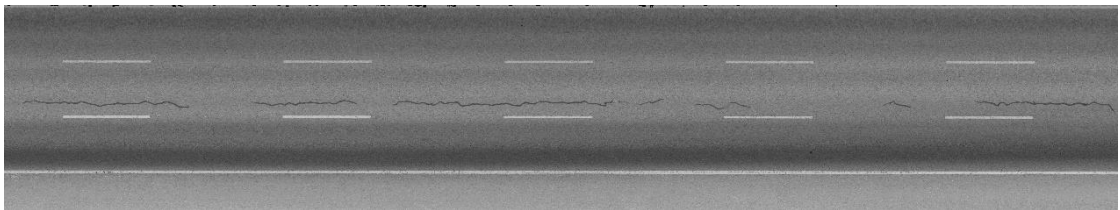

图 3.4 内插后的高分辨率道路面特征图像

# 第四章 道路破损检测算法

本章着重介绍本文所使用的道路破损检测算法及其原理，依照道路破损检测流程顺序分为三个部分进行介绍，分别为图像预处理、二值化分割和道路破损增强。

## 4.1 图像预处理

如前所述，利用车载激光雷达获得的高密度激光点云以及基于其内插得到的道路面特征图像，通常情况下会包含许多噪声以及道路面上的其它干扰物，且车载激光点云的强度受到其采集条件的影响。对于同一观测物体，扫描距离越远，扫描入射角越大，则其反射强度就越小，这便会使得内插得到的二维特征图像的亮度也不均一，所以在道路破损检测前需要进行图像预处理，来减弱甚至消除这些干扰因素对检测结果的影响。具体来说，本文采用了中值滤波、形态学底帽变换来进行图像预处理。

### 4.1.1 中值滤波

中值滤波是一种数字图像处理中常用的非线性平滑技术，它是基于排序统计理论的一种能有效抑制噪声的处理技术。中值滤波的基本原理确定一个像素点的邻域，然后将该像素点和其邻域内像素点的灰度值进行排序处理，将该序列中间位置的灰度值作为该像素点滤波后的灰度值，这样依次遍历整个图像。中值滤波的输出为：

$$g(x, y) = med\{f(x-k, y-l), (k, l \in W)\} \tag{4.1}$$

其中，f(x,y)为滤波前图像，g(x,y)为滤波后图像，W为邻域，邻域可以是不同形状，如线状、方形、圆形、十字形等。

中值滤波可以去除一些孤立的噪声点，减弱道路面上孤立的黑斑对检测结果造成的影响。此外，中值滤波还具有能够保护信号边缘，使之不被模糊的良好特性。所以使用其进行滤波时，不会对道路破损的边缘部分产生太大影响。

本文采用了 3*3 的正方形邻域进行图像平滑，平滑结果如下图所示：

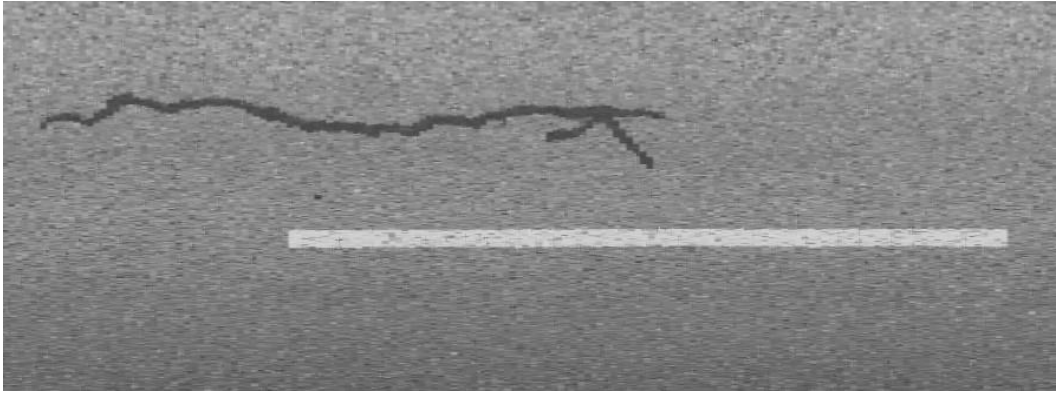

图 4.1 中值滤波前的道路面特征图像

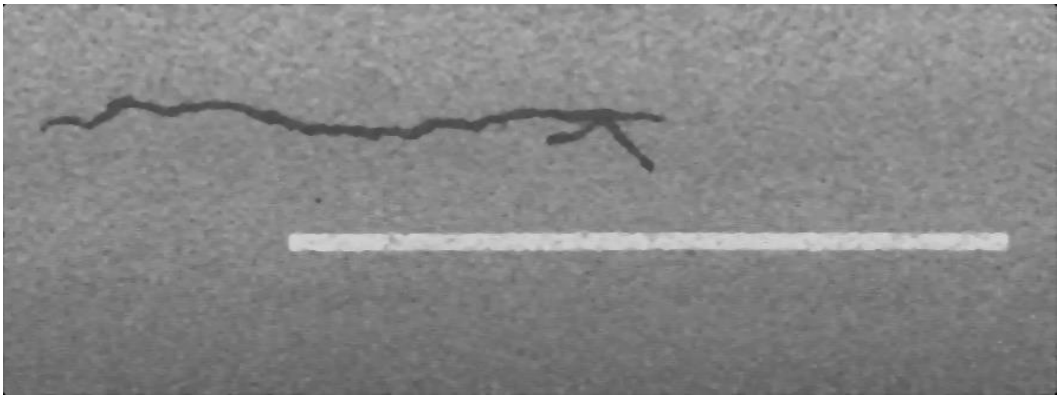

图 4.2 中值滤波后的道路面特征图像

### 4.1.2 底帽变换

底帽变换是基于数学形态学的一种图像处理方法。数学形态学是一种基于集合论、拓扑论和网格代数的空间结构理论，其首要任务是分析目标的形状和结构。数学形态学的主要目的是使用一个特定的结构元素去提取图像上的所对应的结构与形状，以达到对图像分析和识别的目的。数学形态学的基本运算有两个：腐蚀、膨胀。我们可以通过这两种基本运算推到组合出各种其它的数学形态学算法来对图形进行处理，以满足我们的需要，底帽变换就是由此而来的。

二值形态学是数学形态学的基础，其用于处理二值图像。二值形态学将二值图像看成集合，用结构元素对图像集合进行探查。形态学算子表达了结构元素与被提取的图像之间的关系，结构元素决定了提取的结构信息。二值形态学可以被用来对二值图像进行边缘提取、骨架抽取、细化、图像分割、形状分析。对二值形态学进行扩展，以极小值和极大值分别代替交和并运算，用于处理灰度图像，就形成了灰度形态学。

在灰度形态学中，分别用f(x,y)和b(x,y)表示目标图像和结构元素，其中(x,y)表示图像中像素点的坐标

- 腐蚀操作：

  灰度形态学中腐蚀操作的含义为计算该图像点局部范围内各点与结构元素中对应点灰度之差，并选取其中的最小值作为该点的腐蚀结果。

  $$f\Theta b(s,t) = \min\{f(s+x,t+y) + b(x,y) | (s+x,t+y) \in D_f; (x,y) \in D_b\} \quad (4.2)$$

  其中$D_f$和$D_b$分别是f和b的定义域。

- 膨胀操作：

  灰度形态学中膨胀操作的含义为计算该图像点局部范围内各点与结构元素中对应点的灰度之和，并选取其中的最大值作为该点的膨胀结果。

  $$f \oplus b(s,t) = \max\{f(s-x,t-y) + b(x,y) | (s-x,t-y) \in D_f; (x,y) \in D_b\} \quad (4.3)$$

  其中$D_f$和$D_b$分别是f和b的定义域。

- 开操作：

  使用结构元素 B 对集合 A 进行开操作，表示为B∘A，定义为用结构元素 B 对集合 A 进行腐蚀操作，然后用 B 对结果进行膨胀操作。

  $$A \circ B = (A\Theta B) \oplus B \quad (4.4)$$

- 闭操作：

  使用结果元素 B 对集合 A 的闭操作，表示为B●A ，定义为用结构元素 B 对集合 A 进行膨胀操作，然后用 B 对结果进行腐蚀操作。

  $$A \bullet B = (A \oplus B)\Theta B \quad (4.5)$$

- 底帽变换：

  底帽变换定义为：

  $$B_{hat}(f) = (f \bullet b) - f \quad (4.6)$$

  其中B为经底帽变换后的图像，f为底帽变换前的图像，b为结构元素。

  由底帽变换的定义可知，底帽变换可以分解为三个步骤，即先确定一个结构元素，使用该结构元素先对原图进行膨胀，接着在膨胀的基础上，使用相同的结构元素进行腐蚀，上述两步操作即上文介绍的闭操作。最后将其减去原图，得到最终结果。

  就道路面二维特征图像而言，在对图像进行膨胀时，任何宽度小于结构元素

的道路破损都会被填充。膨胀结果如图 4.3 所示。接着，在膨胀后的基础上进行腐蚀，经图像腐蚀后，道路破损以外的区域都会被重构，反映在图像中，则为道路标线的长度和宽度与膨胀前图像中道路标线的长宽相等。腐蚀结果如图 4.4 所示。由于腐蚀后的图像重构了道路破损之外的区域，所以将腐蚀的结果减去原图就可以去除道路标线等道路面干扰物，同时也能够使道路面特征图像亮度更均衡，从而方便检测道路破损。底帽变换的结果如图 4.5 所示，可以看到道路破损在底帽变换中被增强了，道路面特征图像的亮度和对比度更加均一，同时道路标线等干扰物也被成功去除。最后，再将底帽变换后的图像进行灰度反转，得到如图 4.6 的图像，使其有利于使用局部阈值进行道路破损检测。

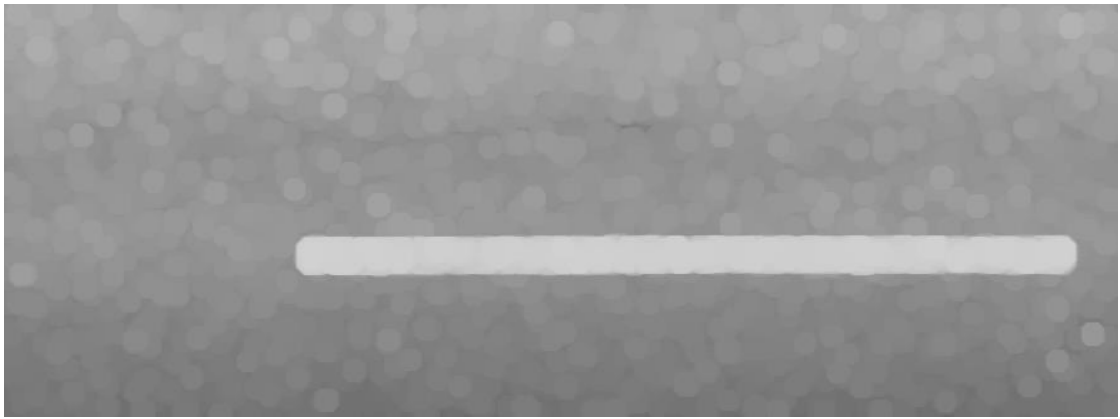

图 4.3 膨胀后的道路面特征图像

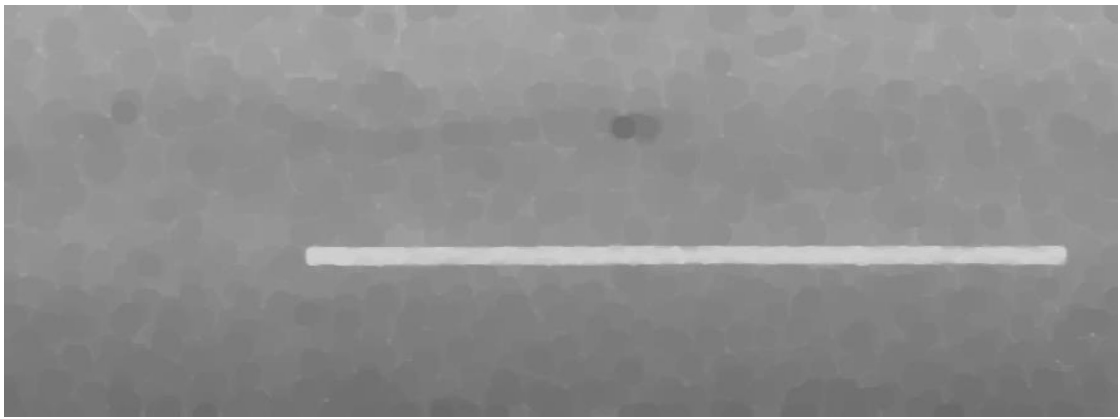

图 4.4 腐蚀后的道路面特征图像

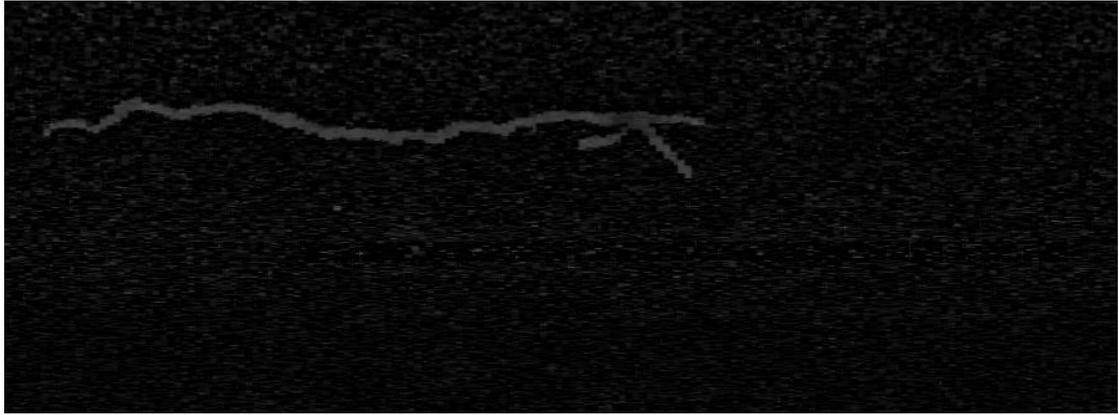

图 4.5 底帽变换后的道路特征图像

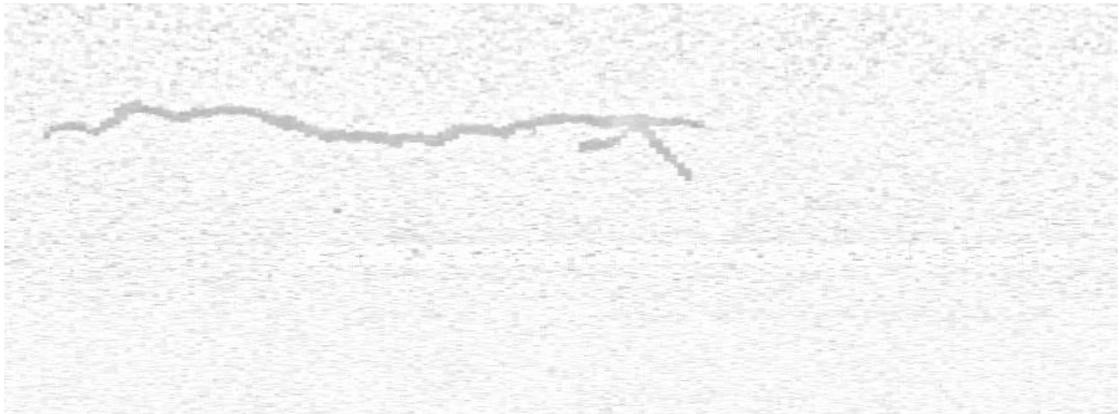

图 4.6 灰度反转后的底帽变换图像

为了更清晰地展现底帽变换中图像灰度的变化过程，现提取道路面特征图像第 300 列的灰度剖面，及其在底帽变换中每一步的变化，将其展示如下：

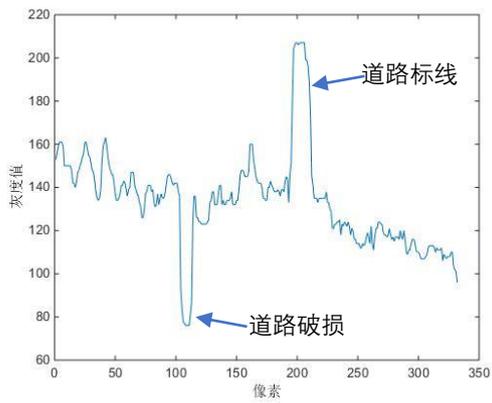

(a)

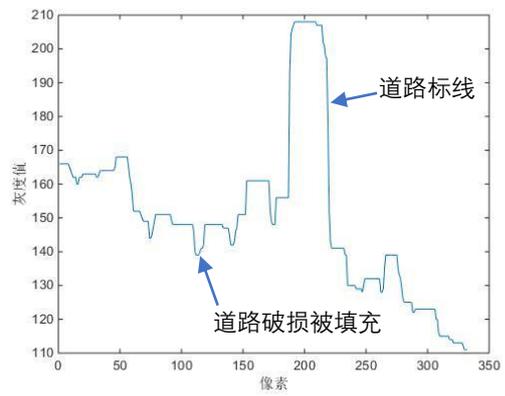

(b)

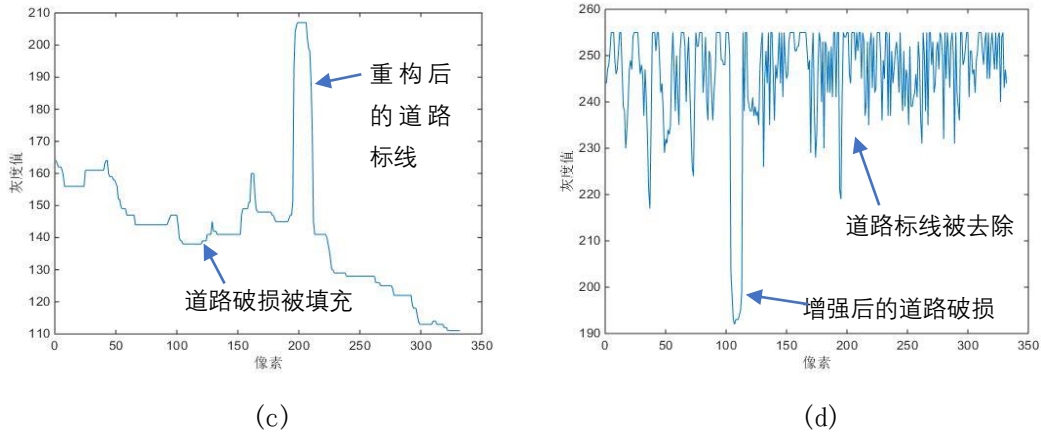

图 4.7 底帽变换中图像的灰度剖面图

(a)为底帽变换前图像的灰度剖面图，(b)为使用半径为 15 的圆形结构元素进行膨胀后的图像灰度剖面图，(c)为使用相同结构元素进行腐蚀后的图像灰度剖面图，(d)为经过底帽变换和灰度反转后的灰度剖面图，可以看出道路破损被很好地增强，道路标线等干扰物也被去除。

## 4.2 二值化分割

图像二值化分割就是将图像上像素点的灰度值设为 0 或 255，即将 256 个亮度等级的灰度图像通过适当的阈值选取而获得仍然可以反映图像整体和局部特征的二值化图像。二值化后的图像分为前景和背景两部分，其灰度值分别为 0 和 255。二值化分割的重点是确定阈值，目前全局阈值和局部自适应阈值是两种主流的方法。

全局阈值是确定一个单一的阈值，对整幅图像都使用一个统一的阈值来进行二值化。大津法（OTSU 法）是全局阈值二值化的代表方法之一。全局阈值的计算速度快，对普通的图像分割的效果较好，但是当图像较为复杂，例如存在许多背景噪声或图像的亮度和对比度不均时，就会有许多像素点不能够被合理地分为前景和背景。

局部自适应阈值则是根据像素邻域像素的灰度值分布来确定该像素位置上的二值化阈值，所以在局部自适应阈值中，每个像素位置处的二值化阈值不是固定不变的，而是由其周围邻域像素的分布来决定。因此灰度较高图像区域的二值化阈值通常会较高，而灰度较低图像区域的二值化阈值则会相适应地变小。Singh T.R.et al[12]，Bernsen et al[13]，Chow and Kaneko et al[14]，J.

Sauvola[15]提出的方法即属于局部自适应阈值二值化方法。

由于道路面特征图像上道路破损的灰度值比背景的灰度值小很多，所以进行二值化分割可以突出道路破损信息，便于道路破损检测。在本研究中，即使道路面特征图像已经经过了预处理，亮度和对比度不均的情况已经明显减弱，但其还是存在。同时，图像上还存在着许多噪声。鉴于这种情况，使用局部自适应阈值能更好地帮助我们从图像中提取道路破损。本文使用 Singh T. R. 等人提出的局部自适应阈值的二值化方法进行二值化分割。

### 4.2.1 Singh T. R. 等人提出的局部自适应阈值的二值化方法

Singh T. R. 等人于 2011 年提出了基于局部平均值和平均偏差的局部自适应阈值二值化方法。[12]像素$I(x,y)$的阈值$T(x,y)$是根据局部平均值$m(x,y)$和平均偏差$\partial(x,y)$由下列公式所计算的。

$$T(x,y) = m(x,y)[1 + k\left(\frac{\partial(x,y)}{1-\partial(x,y)} - 1\right)] \tag{4.7}$$

其中，$m(x,y)$是以$(x,y)$为中心，大小为$w \times w$的邻域窗口内像素灰度的平均值，即局部平均值。平均偏差$\partial(x,y) = I(x,y) - m(x,y)$。$k$为偏差值（bias），其取值范围为$[0,1]$，由使用者设定。该值可以控制计算出的阈值。$k$越小，阈值越大；$k$越大，阈值越小。

由于 J. Sauvola[15]等人提出的方法需要计算每个像素点邻域窗口内像素灰度的标准差，而该方法不需要，所以该方法在计算时间上要比其它局部自适应阈值二值化方法短许多。

同时，为了更快地计算出局部平均值，Singh T. R. 等人引入了积分图的概念，这使得整个方法的计算时间进一步缩短。

对于一幅灰度的图像，积分图像中的任意一点$(x,y)$的值是指从图像的左上角到这个点的所构成的矩形区域内所有的点的灰度值之和，即：

$$g(x,y) = \sum_{i=1}^{x} \sum_{j=1}^{y} I(i,j) \tag{4.8}$$

其中，$g(x,y)$为积分图中$(x,y)$处的值，$I(x,y)$为原图$(x,y)$处的灰度值。

图 4.8 直观地展示了计算积分图的步骤。

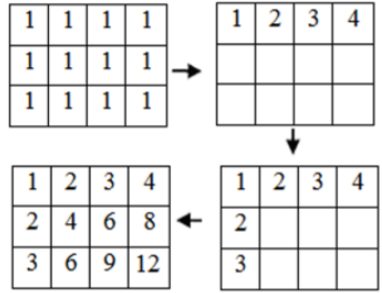

图 4.8 计算积分图的步骤

以像素点$(x, y)$为中心的邻域窗口的局部灰度和$s(x, y)$即为：

$$s(x, y) = \sum_{i=x-c}^{x+c} \sum_{j=y-c}^{y+c} I(i, j) \qquad (4.9)$$

其中，$c = \frac{w-1}{2}$，$w$为窗口大小，为奇数。

若使用积分图计算，则：

$$s(x, y) = [g(x+d-1, y+d-1) + g(x-d, y-d)]$$
$$- [g(x-d, y+d-1) + g(x+d-1)]$$

$$(4.10)$$

其中，$d = round(\frac{w-1}{2})$

因此，以像素点$(x, y)$为中心，窗口大小为$w$的邻域窗口的局部平均值$m(x, y)$即为：

$$m(x, y) = \frac{s(x, y)}{w^2} \qquad (4.11)$$

由上述可见，在使用了积分图后，计算时间不再与窗口大小有关，而通常局部平均值的计算时间与邻域窗口大小相关，因此该方法使得计算时间大大缩短。若图像大小为$n \times n$，邻域窗口大小为$w \times w$，使用传统方法计算$m(x, y)$的时间复杂度为$O(n^2 w^2)$，而该方法的时间复杂度为$O(n^2)$。

### 4.2.2 二值化分割算法的参数调试

在 Singh T.R. 等人提出的方法中，偏差值$k$与窗口大小$w$是两个需要设置的重要参数。

偏差值$k$与阈值成反比。当$k$的数值过小时，则阈值过大，这样会导致道路面上其它干扰物被划分到前景上，造成许多噪声。图 4.9 当$k = 0.01$，$w = 51$时二值化分割后的结果，可见分割后的图像除了有道路破损以外，还有很多噪声。

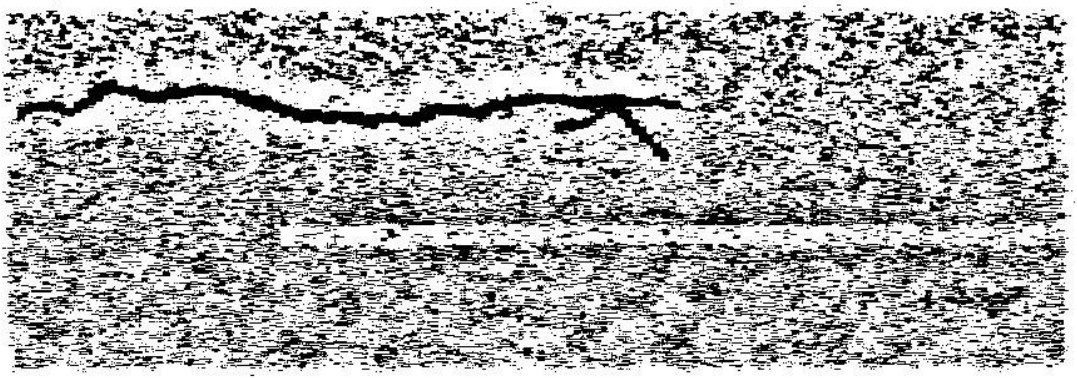

图 4.9 $k = 0.01$，$w = 51$时二值化分割后的结果

当$k$的数值过大时，则阈值过小，虽然这样可以减少被划分到前景的噪声，但同样会导致边缘处灰度较高的道路破损区域不能被划分到前景中。图 4.10 为$k = 0.1$，$w = 51$时二值化分割后的结果，可见到道路破损的边缘区域和灰度值较高的区域没有被划分到前景中。

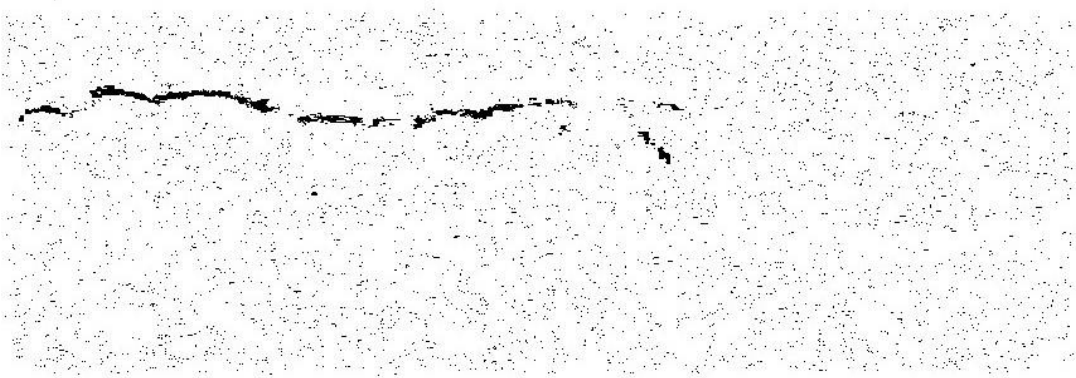

图 4.10 $k = 0.1$，$w = 51$时二值化分割后的结果

邻域窗口大小$w$也与阈值息息相关，因为窗口的大小决定了计算阈值时的所考虑的信息量。在数据集中，道路破损一般的宽度为 20 个像素左右。若邻域窗口大小$w$小于 20，则道路破损中间部分的像素的邻域内会恰好全部是道路破损，换言之，该邻域窗口将不会获得道路破损之外区域的信息，这会导致在计算阈值时失去参照，使得到的数值没有意义。图 4.11 为$k = 0.06$，$w = 15$时，二值化分割后的结果，可见在道路破损宽度较大的区域，计算得到的阈值将该区域内的像素划分为了背景。

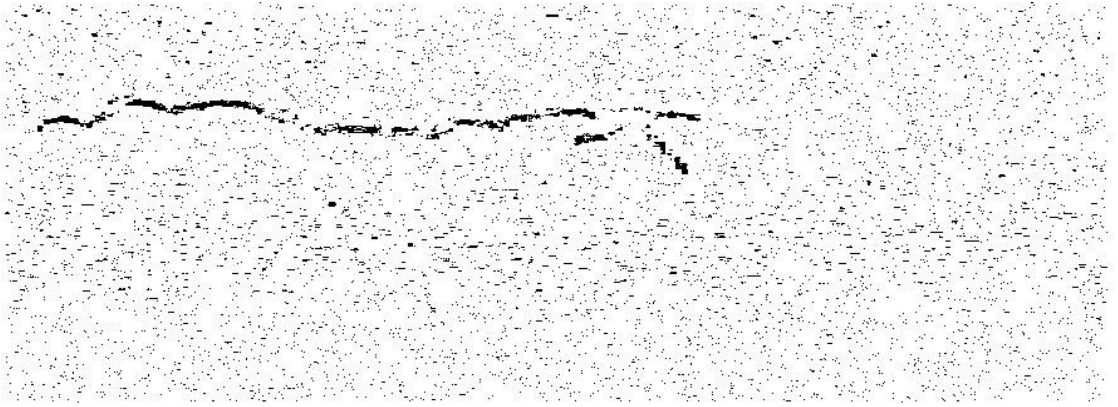

图 4.11 $k = 0.06$，$w = 15$时二值化分割后的结果

同样，邻域窗口也不能太大，如果太大的话，则二值化分割效果就会与全局阈值的分割效果趋近，从而失去局部自适应阈值的意义。

### 4.2.3 二值化分割的结果

本文使用了 Singh T. R. 等人提出的方法进行二值化分割。经过试验，发现当偏差值$k$为 0.06，邻域窗口大小$w$为 51 时的分割效果最为理想。图 4.12 为道路面特征图像原图，图 4.13 为经过底帽变换后的图像，图 4.14 为基于图 4.13，且使用上述参数进行二值化分割后的图像。由此可见，使用该方法进行二值化分割效果较为理想。图 4.15 是使用全局阈值二值化分割方法的代表——大津法（OTSU），同样基于图 4.13 进行二值化后的图像，可见使用局部阈值进行二值化分割的效果明显更优。

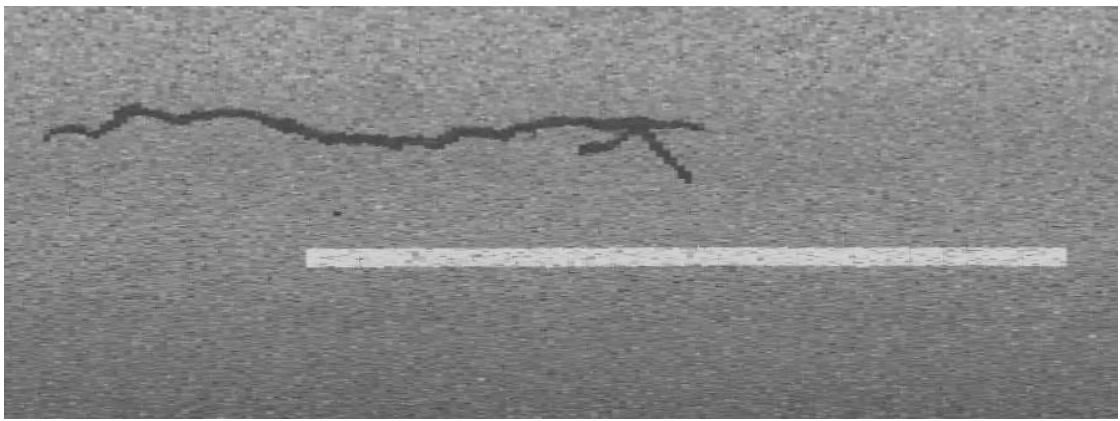

图 4.12 道路面特征图像原图

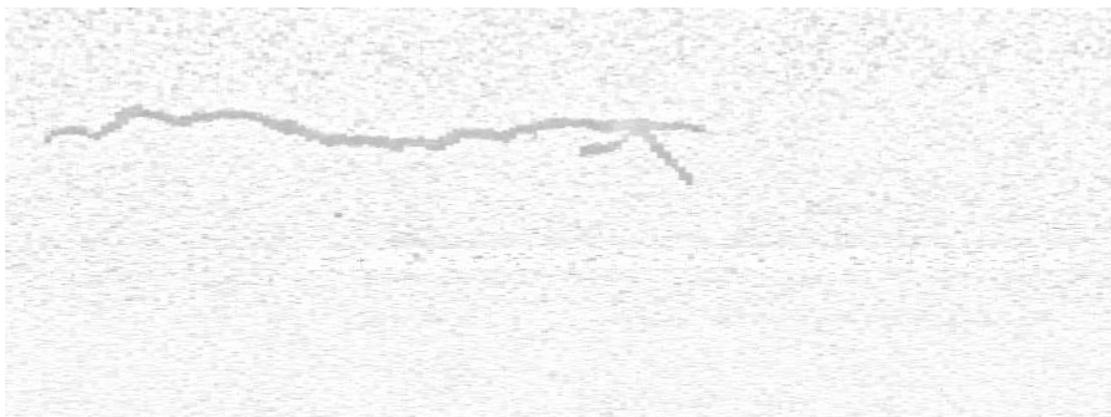

图 4.13 经过底帽变换后的图像

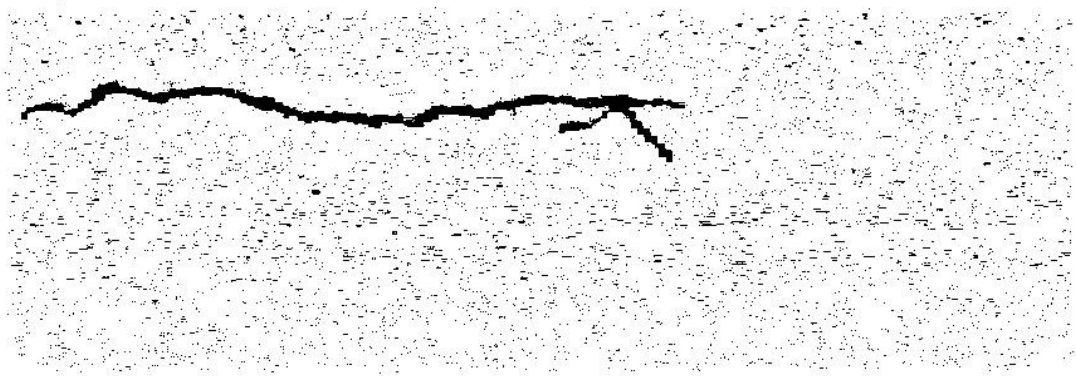

图 4.14　$k = 0.06$，$w = 51$ 时二值化分割后的结果

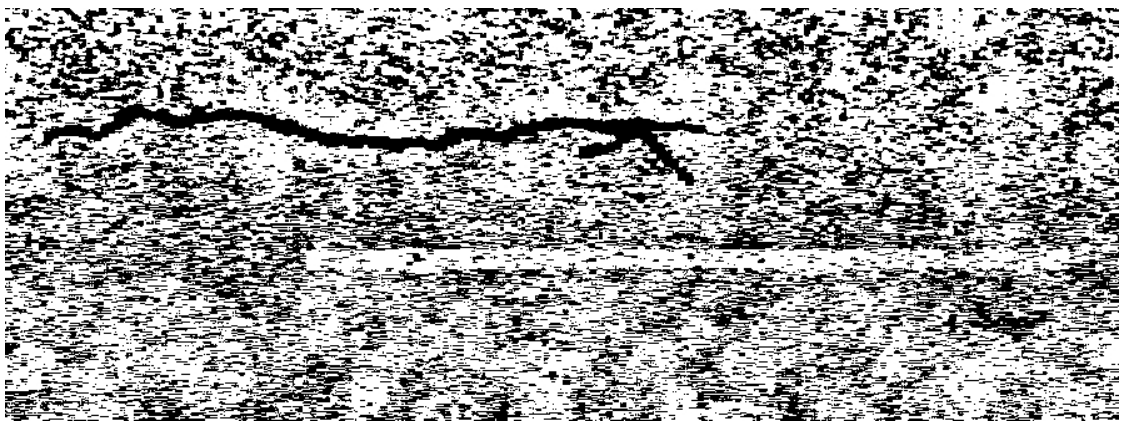

图 4.15 使用大津法进行二值化分割后的结果

## 4.3 道路破损增强

经局部自适应阈值进行二值化分割后的图像像素可分为前景像素和背景像素。理想情况下，前景像素应仅仅包含道路破损信息，但现实情况是前景像素不仅仅包含道路破损，还包含有许多噪声点，我们称之为"假破损"，这对道路破损检测造成了很大的影响，所以我们需要将这些假破损变成背景像素。

假破损是由许多因素造成的，例如车载激光点云扫描系统作业时的采集环境，道路面的属性和仪器的影响等[16]。这些因素导致了部分道路面完好区域像素的灰度值较低，与道路破损处像素的灰度值相似，在二值化分割时，这些像素被错误地分为了道路破损，造成了假破损。道路破损增强的目的就是为了减小假破损的情况。本文使用基于多尺度迭代的张量投票方法进行道路破损增强。

### 4.3.1 张量投票算法

张量投票是计算机视觉里一种重要的感知重组方法，主要解决与 2-D，3-D 及高位的视觉相关的问题。该算法在 Gestalt 心理学基础上，通过特征的张量表示和非线性投票从图像中推断显著性结构，例如点、线、面等。[17]投票是指某一像素点对另一像素点的影响程度，该程度基于 Gestalt 心理学中的接近性、连续性和共势原则，由张量来进行刻画。张量的大小和形状由其特征值（$\lambda 1$，$\lambda 2$ 且 $\lambda 1 \geq \lambda 2 \geq 0$）和特征向量（$\hat{e}_1$，$\hat{e}_2$）所表示。在二维空间中，输入的点可以被任意二阶对称半正定的张量所表示，该张量可以被分解为：

$$T = \lambda_1 \hat{e}_1 \hat{e}_1^T + \lambda_2 \hat{e}_2 \hat{e}_2^T$$
$$= (\lambda_1 - \lambda_2)\hat{e}_1 \hat{e}_1^T + \lambda_2(\hat{e}_1 \hat{e}_1^T + \hat{e}_2 \hat{e}_2^T) \qquad (4.12)$$

其中，$\lambda_1$ 和 $\lambda_2$ 为特征值（$\lambda_1 \geqslant \lambda_2$），$\hat{e}_1$ 和 $\hat{e}_2$ 是其对应的特征向量。在几何中，可以将二维张量视为椭圆，椭圆长短半轴的大小由特征值来确定，而其方向则由特征向量来确定。张量的形状决定了像素点所在区域的结构（例如曲线），而张量的大小则决定了该结构的显著性强度。在上式中，$(\lambda_1 - \lambda_2)\hat{e}_1 \hat{e}_1^T$ 被称为棒张量，棒张量表示该像素点位于曲线上，即该点为曲线元素，$\hat{e}_1$ 表示曲线在该点的法向，显著性强度 $\lambda_1 - \lambda_2$ 则体现了该点是曲线元素的概率，$\lambda_1 - \lambda_2$ 越大，则该点越有可能是曲线元素。$\lambda_2(\hat{e}_1 \hat{e}_1^T + \hat{e}_2 \hat{e}_2^T)$ 被称为球张量，球张量的特征向量方向均一，表明该像素点为点状元素，可能位于曲线交叉处、边缘等区域，也可能为孤立点，$\lambda_2$ 为其显著性强度指标。因此，不同的椭圆形状就定义了不同的张量类型，即棒张量和球张量。

在张量投票中，输入的像素点首先被编码为张量，若像素的法向量未知，则将其编码为球张量，即：$T = \begin{bmatrix} 1 & 0 \\ 0 & 1 \end{bmatrix}$，其特征值均为 1，；若像素的法向量已知（$n_1$，$n_2$），则将其编码为棒张量，其特征值为 $\lambda_1 = 1$，$\lambda_2 = 0$，即：$T =$

$\begin{bmatrix} n_1^2 & n_1 n_2 \\ n_1 n_2 & n_2^2 \end{bmatrix}$。

在编码完成之后，这些张量的几何信息就会通过张量投票来传递给其邻域的张量。假设$O$点为投票点，位于原点，向$P$点（接收点）进行投票，它们的法向量分别为$N_C$、 $N_P$，从$O$点到$P$点的弧线$L$就是基于平滑性和临近性原则定义的最短路径。 $\theta$为以 $C$ 为圆心的密切圆在$O$处的切线与$OP$之间的夹角，$OP$的弧长$s = \frac{\theta l}{\sin(\theta)}$，$OP$的曲率$\kappa = \frac{2\sin(\theta)}{l}$。则其投票方式为：

$$\mathbf{S}_{\mathrm{SO}}(l, \theta, \sigma) = DF(s, \kappa, \sigma) \begin{bmatrix} -\sin(2\theta) \\ \cos(2\theta) \end{bmatrix} \begin{bmatrix} -\sin(2\theta) & \cos(2\theta) \end{bmatrix}$$

(4.13)

其中， 衰减函数$DF(s, k, \sigma) = e^{-\left(\frac{s^2 + ck^2}{\sigma^2}\right)}$，$c$控制衰减程度，其定义为$c = \frac{-16\log(0.1) \times (\sigma-1)}{\pi^2}$，$\sigma$为尺度参数，决定投票区域大小，是张量投票中唯一可以人为设定的参数。接收点$P$点所接受到的大小是由衰减函数$DF(s, k, \sigma)$所确定的。张量投票的示意图如图4.16所示。

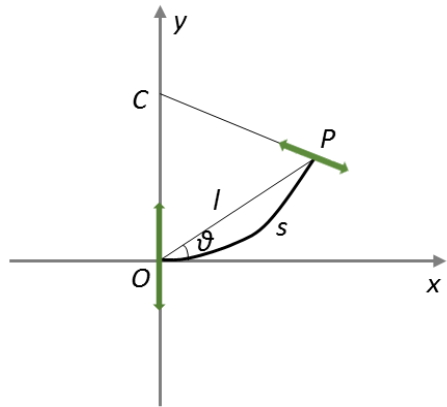

图4.16 张量投票示意图

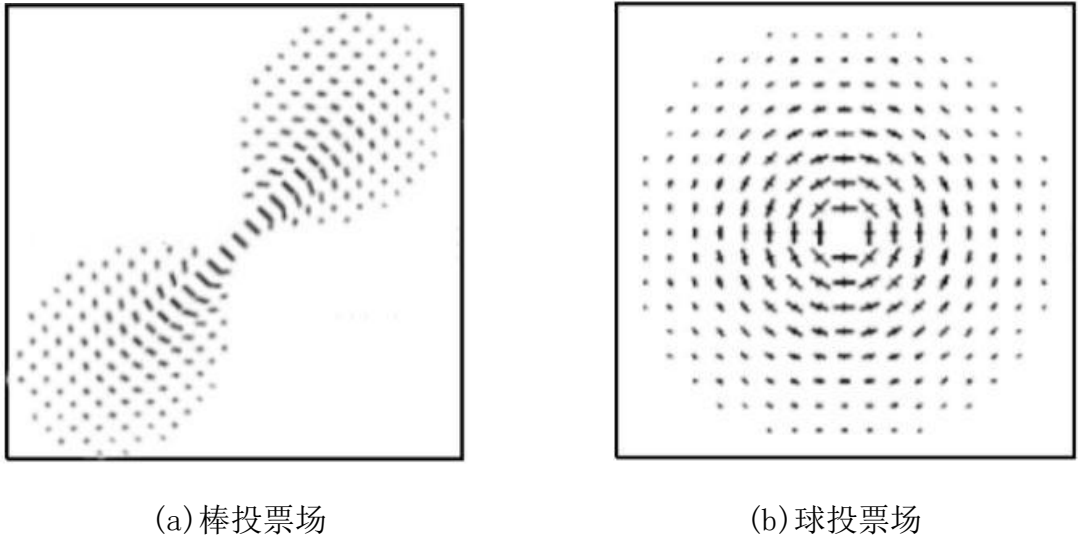

(a)棒投票场          (b)球投票场

图 4.17 张量投票场

    为了加快投票速度，在投票的时候不直接计算每一个张量的投票值，而是通过预先建立棒投票场（见图4.17(a)）和球投票场（见图4.17(b)），棒投票场和球投票场内储存着棒张量和球张量在不同投票距离和投票角度邻域内事先计算好的张量投票[11]，在投票的时候只需通过与待处理张量卷积便可得到值。使用球投票场的张量投票为球张量投票，使用棒张量场的张量投票为棒张量投票。尺度参数 $\sigma$ 控制投票场的大小，$\sigma$ 越大，则投票的范围越大，这样可以抑制噪声，但会使得图像上的许多细节被平滑掉。$\sigma$ 越小，则投票的范围越小，这样可以保存更多细节，但会导致算法对噪声的鲁棒性不强。区别于球张量场，棒张量场的范围被限制在了 $|\theta| \leq 45°$ 的区域内，在这个区域之外，O至P的最光滑路径就不能被张量O和张量P所形成的密切圆所表示。

    每个张量都会收集其邻域内其它张量对其的投票并将投票累积，从而形成一个新的张量，反映其几何结构特征。投票的累积是通过张量相加来完成的，具体而言就是 $2 \times 2$ 的矩阵相加。在投票结束后，每一个像素点即可获得一个新的张量。再利用前述方法进行张量分解，即可得球张量和棒张量的显著性强度图 $\lambda_2$ 和 $\lambda_1 - \lambda_2$，分别代表点状元素和线状元素的置信度。通过分析球张量和棒张量的显著性强度，我们能够基本确定像素点是否为噪声点，若 $\lambda_1$、$\lambda_2$ 都非常小，则基本可判定该像素为噪声点，从而将其去除。下表为投票后张量所反映的几何结构特征：

表 4.1

| 元素种类 | 显著性强度 | 法线方向 | 切线方向 |
|---|---|---|---|
| 线 | $\lambda_1 - \lambda_2$ | $\hat{e}_1$ | $\hat{e}_2$ |
| 点 | $\lambda_2$ | 无 | 无 |

此外，张量投票分为稀疏张量投票和密集张量投票，稀疏张量投票是只对邻域内的张量进行投票，而密集张量投票是对邻域内的所有点进行投票。

### 4.3.2 基于多尺度迭代的张量投票的道路破损增强

如上文所述，尺度参数 $\sigma$ 控制了投票场的大小，大投票场会使得图像上的噪声被去除，但同时会造成道路破损的细节被平滑掉（见图 4.18(a)）；小投票场则相反，它会保留道路破损的边缘细节部分，但也会造成图像噪声不能被很好地去除（见图 4.18(b)）。

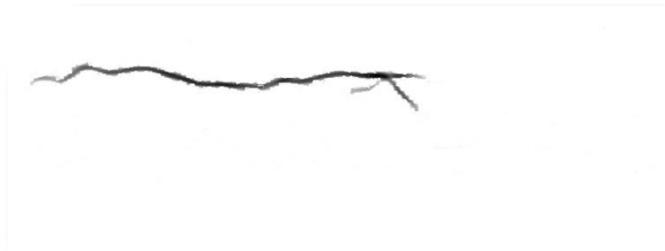

(a) $\sigma$ =20 时的棒张量显著性强度图

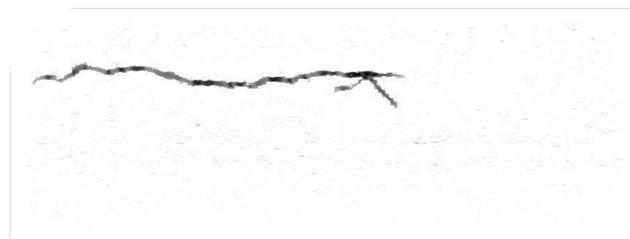

(b) $\sigma$ =2 时的棒张量显著性强度图

图 4.18 棒张量显著性强度图

鉴于此情况，我们提出了多尺度迭代的张量投票算法来解决这一问题。图

4.25 为该算法的流程图。这里我们将一次球张量投票和一次棒张量投票合称为一次张量投票。

首先，我们对输入图像前景部分的像素进行张量编码，由于这些像素没有方向，所以它们只能被编码为球张量，即 $T = \begin{bmatrix} 1 & 0 \\ 0 & 1 \end{bmatrix}$。在编码完成之后，我们使用尺度参数为 $\sigma_{ball}$ 的球投票场进行第一轮稀疏球张量投票。此时，编码后的张量就会向其邻域内的其它张量进行投票，传递信息。

在球张量投票结束后，原先的像素点通过累加得到的投票得到新的张量，就会获得初始的强度（$\lambda_1$，$\lambda_2$）和方向（$\hat{e}_1$，$\hat{e}_2$）信息。将其进行张量分解，便得到了棒张量显著性强度和球张量显著性强度，分别代表某像素点位于线状结构和点状结构的置信度。这里我们使用阈值 $T_{stick1}$ 对得到的棒张量显著性强度图进行过滤，滤除那些显著性强度小的像素，$\lambda_1 - \lambda_2 < T_{stick1}$ 的像素会被去除，即被设为背景。这样可以使得一部分噪声被去除，以免这些噪声点对其它点的投票影响提取结果，同时加快投票速度。在这里，棒张量显著性强度阈值的设定十分保守，只有那些显著性强度非常低的点才会被去除，这样才能够在去除一部分离散噪声点的同时完整保留道路破损的细节部分。图 4.19 为二值化前的棒张量显著性图，可见处于曲线上的像素点的显著性强度值较高。图 4.20 为二值化后的棒张量显著性图，由于阈值设定较为保守，所以被去除的像素点数量并不是很多。

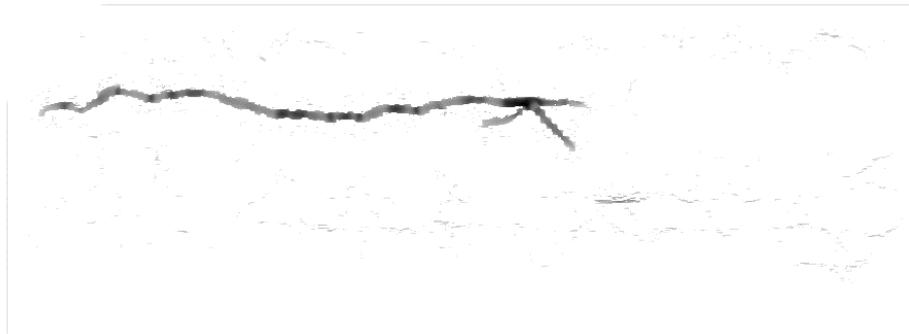

图 4.19 二值化前的棒张量显著性图

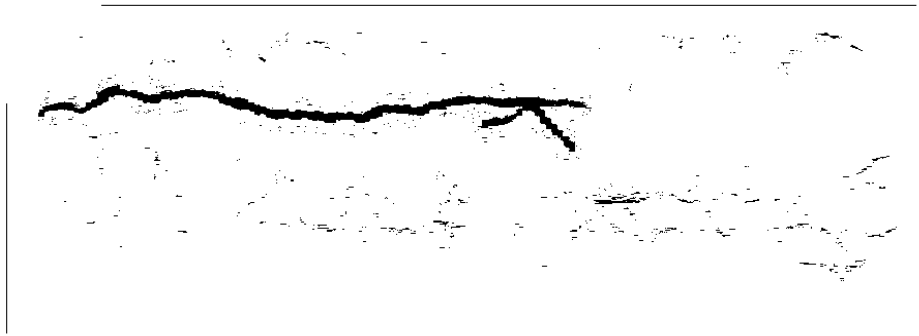

图 4.20 二值化后的棒张量显著性图

由于道路破损的形态多种多样，为了保证提取出的道路破损信息的完整性，我们必须同时考虑棒张量和球张量。因此我们将球张量显著性强度图使用阈值 $T_{ball}$ 进行二值化，并存入变量，与最后一轮投票得到的棒张量显著性强度图合成得到最终结果。图 4.21 为二值化前的球张量显著性强度图，可以发现在道路破损交点处，球张量显著性强度较高，而在棒张量显著性强度图中，这一部分的显著性强度较低，所以二者有互补性，只有同时考虑棒张量显著性强度和球张量显著性强度才能较完整地提取不同形态的道路破损。图 4.22 为二值化后的球张量显著性强度图，道路破损交点处的像素被保留了下来。

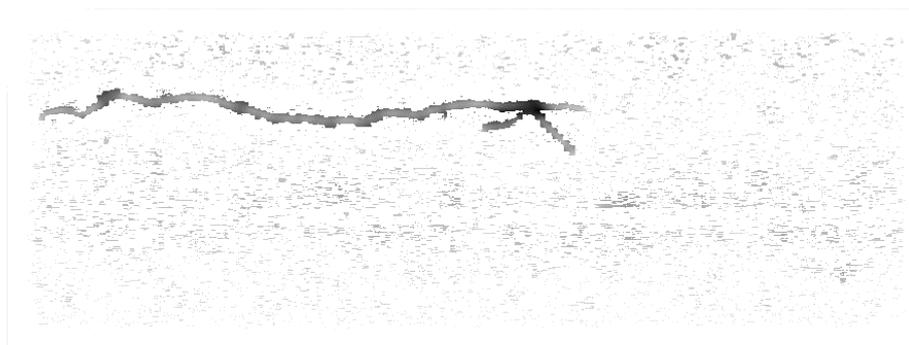

图 4.21 二值化前的球张量显著性强度图

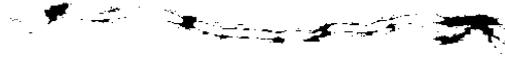

图 4.22 二值化后的球张量显著性强度图

接着，由于在稀疏球张量投票后，张量已经获得了初始的方向信息（$\hat{e}_1$，$\hat{e}_2$），因此接下来我们使用尺度参数为$\sigma_{stick1}$的棒投票场进行稀疏棒张量投票，来对稀疏球张量投票获得的棒张量进行增强，使得棒张量的强度和方向信息更加显著。本轮的稀疏棒张量投票使用小尺度参数，主要目的是为了增强道路破损信息，保护道路破损的细节部分。当然使用小尺度参数进行投票也不可避免地增强了图像上噪声，这一问题可以通过后续轮次的投票来解决。在该轮投票完成后，我们同样通过张量分解，得到棒张量显著性强度图，并使用阈值$T_{stick2}$对得到的显著性强度图进行二值化，滤除显著性强度较低的像素点，并将大于阈值的像素点设为前景。

此外，不同于其它学者提出的方法，本文使用了稀疏棒张量投票而非密集棒张量投票。这是因为本文是以像素级精度提取道路破损的，而非如其他学者，如Guan et al[18]使用形态学细化的方法提取道路破损的中轴线，这就对精度有了更高的要求。密集张量投票会对邻域内的所有点进行投票，若是仅提取道路破损的中轴线，使用密集张量投票可以将不连续的道路破损连接，再使用形态学细化提取出一条完整的中轴线，同时密集张量投票在道路破损周围产生的许多噪声会在形态学细化的过程中被消除；然而若是以像素级精度提去道路破损，这些噪声则会造成提取精度下降，因此本文选用了稀疏张量投票而非密集张量投票。

在此我们已经完成了第一轮张量投票，并得到了二值化后的棒张量显著性强度图和球张量显著性强度图。

第二轮张量投票是基于第一轮张量投票得到的二值化后的棒张量显著性强度图进行的。该轮张量投票的目的是为了减弱图像上噪声的影响，因此选择了较

大的尺度参数进行。投票的过程与第一轮投票相似，同样分别进行稀疏球张量投票和稀疏棒张量投票，具体的投票过程就不再赘述。主要区别就是在棒张量投票中使用了较大的尺度参数$\sigma_{stick2}$和阈值$T_{stick3}$。由于第二轮投票时棒张量显著性强度已经被增强，所以我们使用了比$T_{stick2}$稍大一些的阈值$T_{stick3}$对该轮投票得到的棒张量显著性强度图进行二值化。在经过两次迭代后，道路破损增强就可以取得较为理想的效果。图 4.23 为二值化前的球张量显著性强度图，可见图上的大部分噪声点已经被去除，且道路破损边缘细节部分也被较完整地保留了下来。

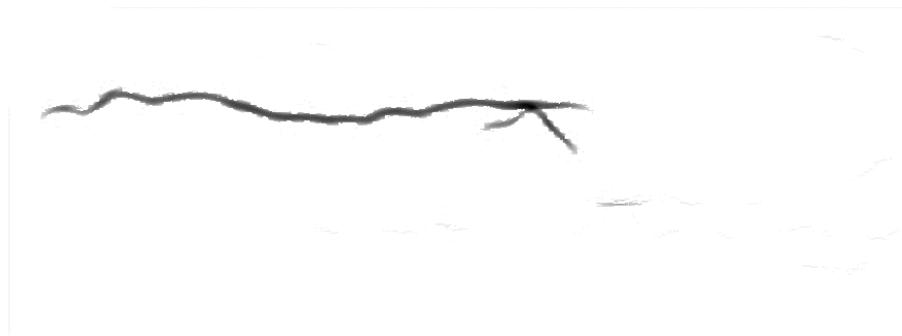

图 4.23  二值化前的球张量显著性强度图

最后，我们将二值化后的棒张量显著性强度图和球张量显著性强度图相加并归一化，使用形态学的方法去除在张量投票过程中造成的道路破损上的毛刺和一些剩余的孤立噪声点，并得到最终增强后的道路破损图像，见图 4.24。有图可见原图上的噪声被去除，道路破损被完整地提取了出来，破损边缘的细节也被很好地保留。

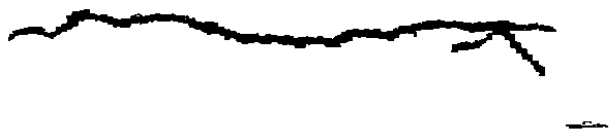

图 4.24  最终提取得到的道路破损

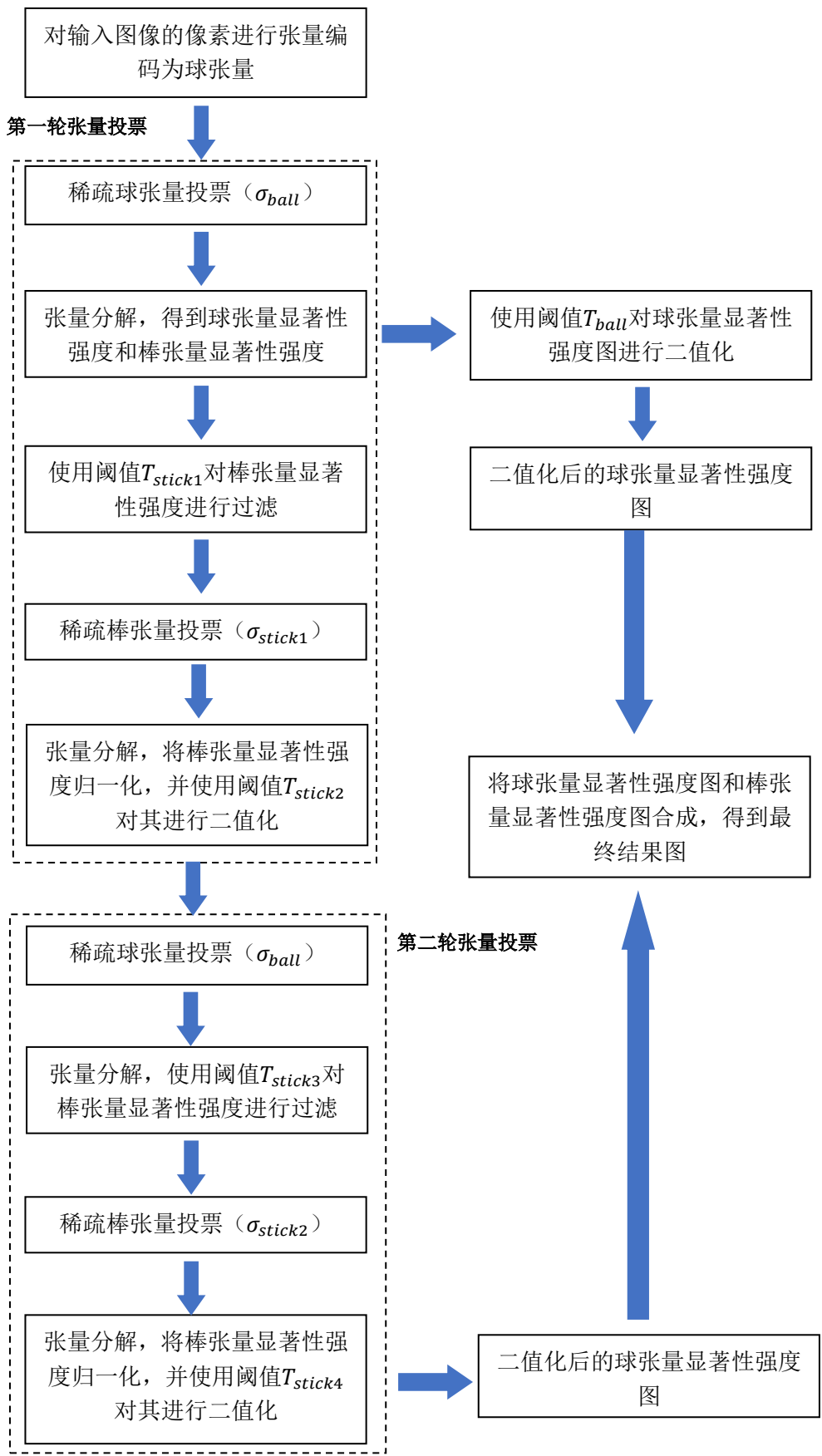

图 4.25 基于多尺度迭代的张量投票算法流程图

# 第五章 算法性能测试与精度评定

## 5.1 算法性能测试

为了更直观的展示裂缝的识别效果，并更方便的对算法性能进行测试，本文使用了 MATLAB R2016a 进行相应的软件编写和测试工作。MATLAB 由美国 MathWorks 公司出品，为一门高级编程语言，它集成了各种高级工具箱，因此操作较为简易方便，其提供的编译器也具备动态编译的高级功能。下表为本文算法在五张不同大小的道路面图像上的运行速度。

表 5.1 五张图像的运行速度

|  | 图 3.3 | 图 3.4 | 图 3.5 | 图 3.6 | 图 3.7 |
|---|---|---|---|---|---|
| 运行时间 | 12.3508s | 7.5231s | 8.0292s | 8.4680s | 24.0281s |
| 图像大小 | 801*455 | 335*667 | 397*606 | 316*573 | 823*1035 |

由此可见，该运算速度尚不能够进行实时处理。图像的大小是影响运算速度的主要因素。在运算中，棒张量投票所占用的时间最长，为性能瓶颈。由于 MATLAB 是解释型语言，所以运算速度较编译语言会慢不少，如果将程序用编译型语言，如 C/C++等语言改写，则运算速度会提高许多。此外，还可考虑将程序在并行环境中运行，以提高运算速度。

## 5.2 精度评定方法

经过上述各个步骤的道路破损信息提取以后，我们需要对提取结果进行统计分析与结果对比，以证明提取结果的可靠性和准确性。本文使用人工标记的道路破损图像作为参考依据来评定本文算法提取出的道路破损的精度。为了客观定量地评价精度，我们采用带搜索区的 Hausdorff 距离作为精度评价方法，并定义了 SM 值作为精度评定指标。

## 5.3 Hausdorff 距离

Hausdorff 距离是根据 Felix Hausdorff 命名的，Hausdorff 距离是指某一集合中的点离另一集合最近点的所有距离最大值，其公式为：

$$H(A,B) = max(h(A,B), h(B,A)) \tag{5.1}$$

其中

$$h(A, B) = max_{a \in A} \, min_{b \in B} \, ||a - b||$$

$$h(B, A) = max_{b \in B} \, min_{a \in A} \, ||b - a||$$

$||a - b||$为集合$A$和集合$B$中点的欧几里得距离，$h(A, B)$为有向 Hausdorff 距离，其刻画了两个点集形状上的差异。$h(A, B)$具体的计算过程可参照下图。

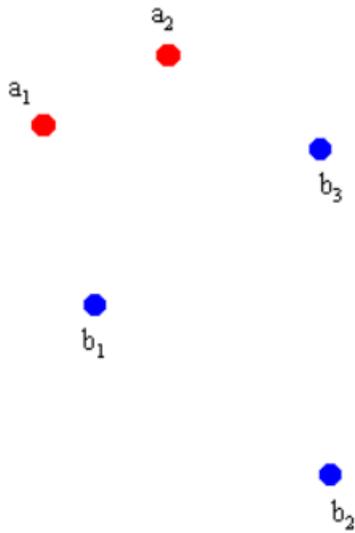

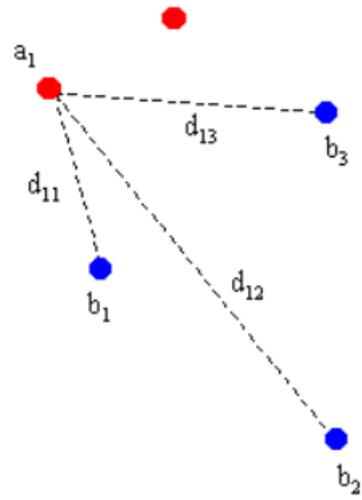

<div align="center">(a)</div>

给定两个点集$A$和$B$，分别用红色和蓝色表示

<div align="center">(b)</div>

计算$a_1$到$b_i$的距离$d_{1i}$，即$d_{11}$, $d_{12}$, $d_{13}$

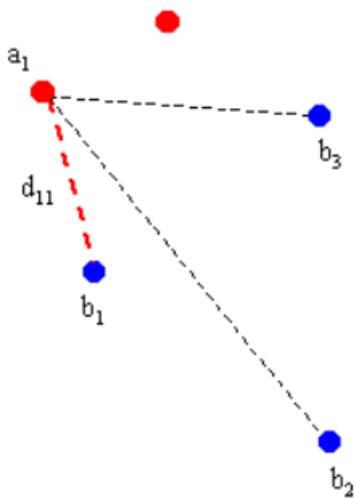

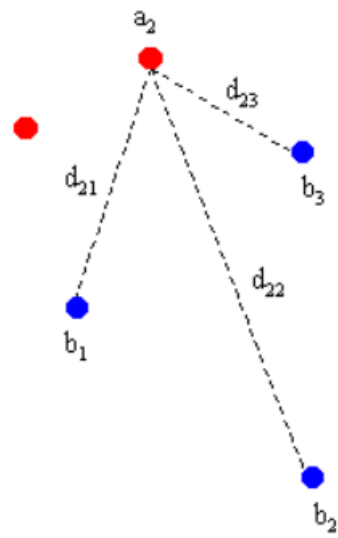

<div align="center">(c)</div>

找到$min(d_{1i})$

<div align="center">(d)</div>

依次类推，计算$a_j$到$b_i$的距离$d_{ji}$

|   (e)   |   (f)   |
|---|---|
| 找到 $min(d_{ji})$ | 找到 $max(min(d_{ji}))$，即为 $h(A, B)$ |

图 5.1 有向 Hausdorff 距离的计算过程

本文使用了带搜索区的 Hausdorff 距离来比较本文提出的算法所提取出的道路破损像素与真实人工标记的道路破损像素之间的差异，从而评价提取精度。$A = \{a_1, a_2, a_3, ..., a_i\}, B = \{b_1, b_2, b_3, ..., b_j\}$ 分别为提取出的道路破损像素点集合和人工标记的道路破损像素点集合，并使用 $H(A, B)$ 计算两者的 Hausdorff 距离。这里我们建立了一个以 $a_j$ 为中心，大小为 $L$ 的搜索区域，在此区域里搜寻距离 $a_j$ 最近的点，这样可以大大加快算法的运行速度。

得到两者的 Hausdorff 距离后，我们定义了精度指标 $SM$ 来评定提取的精度，其定义如下：

$$SM = 100 - \frac{H(A,B)}{L} \times 100 \tag{5.2}$$

$SM$ 值的范围为 0 至 100，$SM$ 值越大，表明精度越高，提取的效果越好。

## 5.4 精度评定结果

针对本文算法的具体应用，在实验中我们选取了五张以沥青路面为背景的含有道路破损的道路面图像，来评定该提取方法的精度。为了确保算法的普适应，本文了选取了不同形态的道路破损图像，包括纵向裂缝、横向裂缝、以及有交叉的龟状裂缝等进行评价。同时，五张道路面破损图像的尺寸大小不同，分别为

801*455，335*667，397*606，316*573，823*1035。

实验结果图如下，其中(a)为道路面破损图像，(b)为人工标记的道路破损参考图像，(c)为本文算法提取出的道路裂缝。

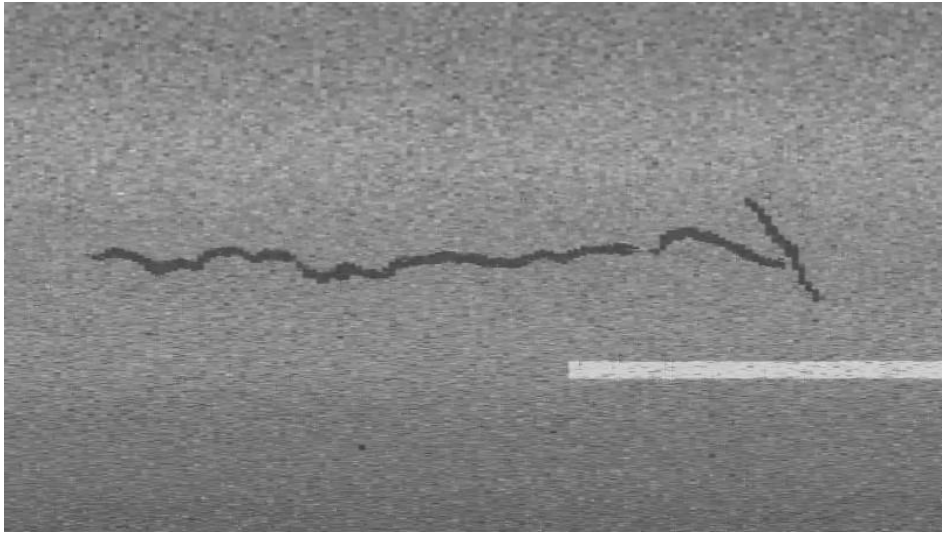

(a)

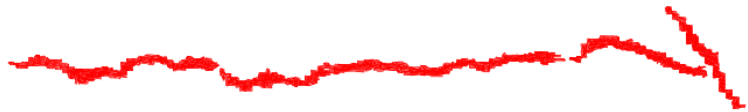

(b)

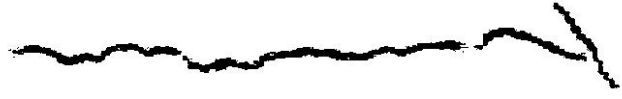

(c)

图 5.2 道路面图像 1

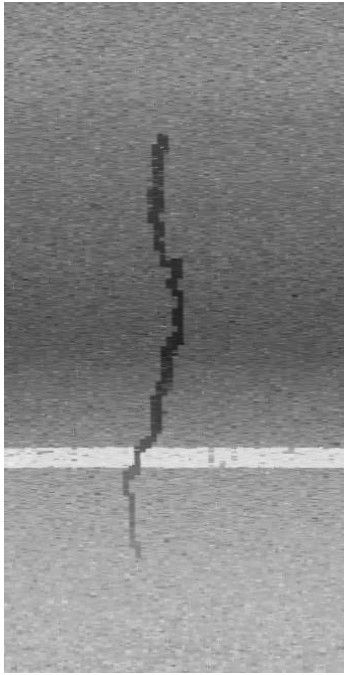

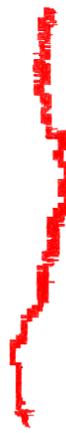

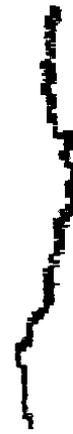

(a)                              (b)                              (c)

图 5.3 道路面图像 2

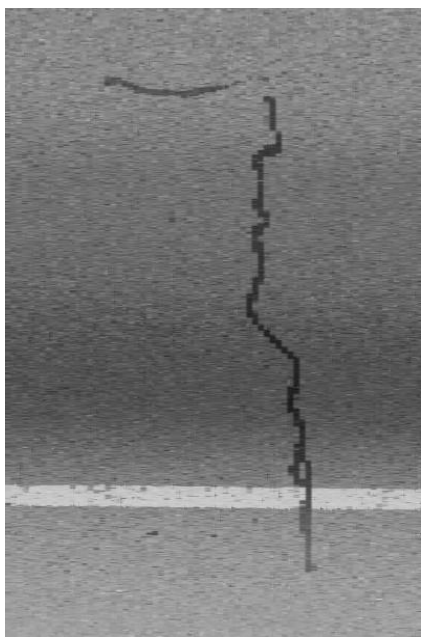 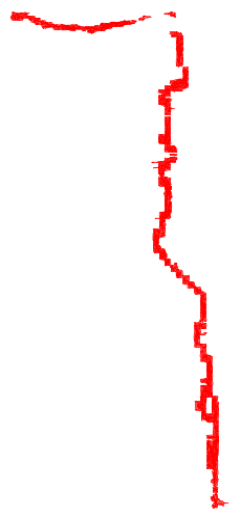

(a)                                           (b)

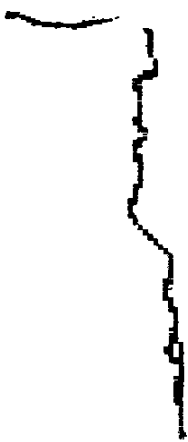

(c)

图 5.4 道路面图像 3

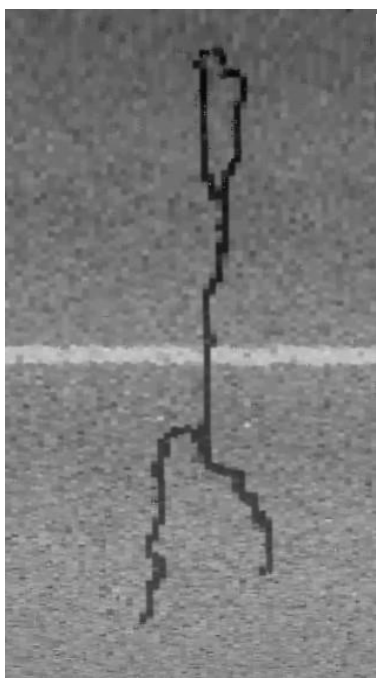 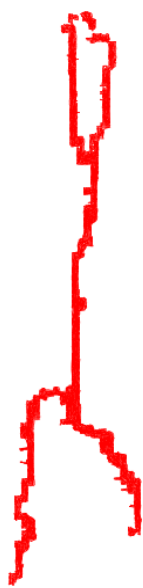 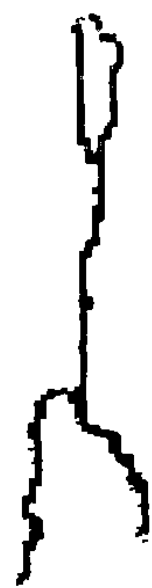

<div style="text-align:center">

(a)  (b)  (c)

图 5.5 道路面图像 4

</div>

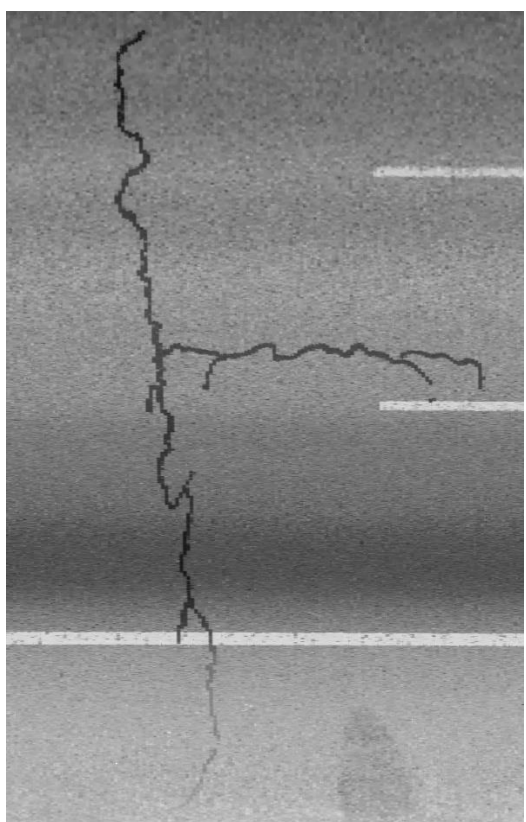 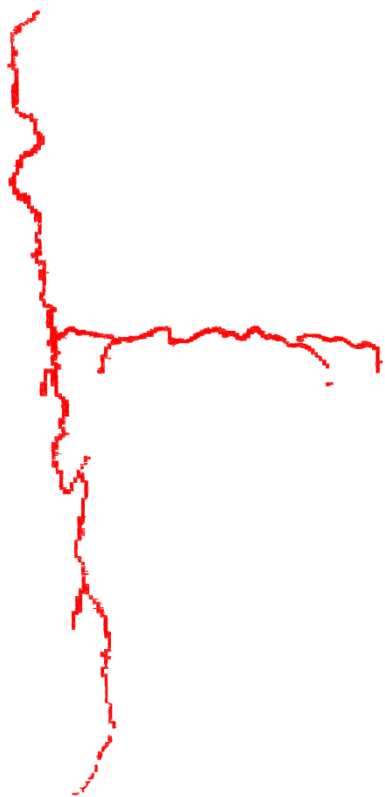

<div style="text-align:center">

(a)  (b)

</div>

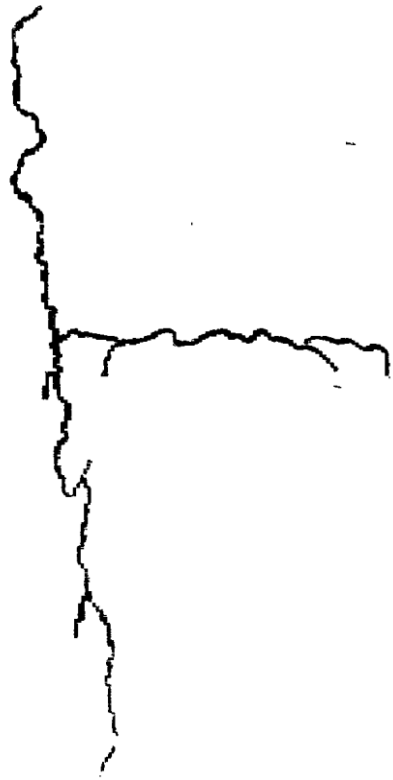

(c)

图 5.6 道路面图像 5

表 5.2 五张图像的 SM 值

|  | 图像 1 | 图像 2 | 图像 3 | 图像 4 | 图像 5 |
|---|---|---|---|---|---|
| SM 值 | 96.5699 | 93.1883 | 93.6437 | 95.0070 | 94.0009 |

可以看出,本文算法提取出的道路破损的 SM 值已达到 95 左右,准确率较高,且较为稳定,这证明了该算法的有效性。同时,该算法对不同类型的道路破损均有较为理想的提取效果,且提取过程不受道路标线等道路面其它干扰物的影响,说明本文提出的算法有较强的鲁棒性。但尽管已经使用形态学底帽变换对图像进行了预处理,图像上一些亮度较高、对比度较低区域处的裂缝仍旧没有被完全提取出来,例如图 5.6 中道路破损的末端,这也造成了其 SM 值偏低。

# 第六章 总结与展望

## 6.1 总结

 我国社会经济已进入飞速发展时期，基础设施建设正如火如荼地进行，特别是公路的建设，目前我国公路通车里程已稳居世界前列。公路是社会经济发展的命脉，只有公路的路况良好才能保证交通运转，因此道路的保养和维护就成为了必不可少的环节。道路路面出现破损是不可避免的，但若不及时对路面病害进行维护，道路的使用寿命将大大减少，还会造成行车安全隐患。为了避免对道路进行反复大面积的维修，及时对道路病害进行检修是唯一的方法。但公路里程长，使用人工进行道路安全检测费时费力，因此使用自动化手段进行道路检测是更为合理的一种方法。

 本文借鉴了前人的道路破损检测技术，融汇创新，提出利用激光点云内插得到的道路面影像，将激光雷达与图像的优点结合起来进行道路破损检测，进行了以下研究工作：

● 了解了道路破损的分类、形成原因以及它们的危害。基于已有的研究，了解不同道路破损提取方法的优点与缺点，熟悉数据集，初步制定算法方案，将其分为数据准备、图像预处理、二值化分割和道路破损增强四步来完成。

● 针对激光点云内插得到的道路面影像的特点，采用了中值滤波进行初步去噪，并利用形态学底帽变换的方法使得图像亮度对比度均一，去除道路标线等干扰物对提取结果的影响。

● 根据道路面内插影像背景噪声较多的特点，采用局部自适应二值化分割算法，对预处理后的道路面图像进行二值化分割，初步提取出道路面破损。

● 针对道路破损的连接性和连续性等特性，使用张量投票算法对道路破损进行增强，并根据实际情况对张量投票算法进行了改进，提出多尺度迭代的张量投票算法并应用在本文的研究中，在保留道路破损边缘细节的同时，去除了图像上的噪声。

● 采用带搜索区的 Hausdorff 距离作为标准进行精度评定，使用 MATLAB 编写算法程序，并进行算法测试。精度结果表明本文提出的方法能够较为准确地从道路面中提取道路破损信息，且基本不受路面干扰物的干扰。

## 6.2 展望

通过实验数据的验证，本文提出的算法较好地实现了从道路面上提取道路破损信息。在已知每像素对应的道路面实际面积时，还可通过本文提出的算法进行道路破损面积估计，这会大大提高道路病害检测的作业效率，降低人工成本。但本算法仍有不足之处，下面列举算法待改进的几个方向：

● 算法运行速度偏慢，特别是在棒张量投票部分。可以对改部分代码进行改进，使其运算速度加快，从而达到实时处理的要求。

● 算法对图像上亮度高、对比度偏低区域的裂缝提取效果仍然有些不太理想。这类区域中的部分裂缝没有被完全提取出来。这和算法的局部自适应阈值部分有着很大的关系，因此还需要对该部分做出改进，更加系统地比较不同参数对提取结果的影响。

● 由于道路破损端点处在张量投票时只能接收到一侧的投票，这导致了端点处获得的投票值偏小，其棒张量和球张量显著性强度值也偏小。这会造成部分破损端点处像素的显著性强度值小于设定的阈值，在提取的过程中被设为背景。可以对张量投票算法进行改进，或者使用其它方法解决这一问题。

# 参考文献

[1] 郭涛. 路面破损信息自动采集技术研究[D]. 武汉理工大学，2006.

[2] 吴芬芳，基于车载激光扫描数据的建筑物特征提取研究[D] 武汉大学,2005

[3] 邢爱萍. 路面使用性能的评价与预测[D]. 北京交通大学，2004.

[4] 李国燕. 基于图像的路面破损识别[D]. 河北工业大学，2009.

[5] 初秀民，沥青路面破损图像识别方法研究[D] 吉林大学,2003

[6] 李刚，贺昱曜，赵妍. 基于大津法和互信息量的路面破损图像自动识别算法[J]. 微电子学与计算机，2009, 26(7):241-243.

[7] Bhagvati C.Skolnick M M,Grivas D A.Gaussian normalization of morphological size distributions for increasing sensitivity to texture variations and its application to pavement distress classification [J].IEEE Computer Vision and Pattern Recognition Conference, Seattle, WA,1994:20-24

[8] Siriphan Jitprasithsiri,Development of a New Digital Pavement Image Processing Algorithm for Unified Crack Index Computation [D].A Dissertation Submitted to the Faculty of the University of Utah, 1997

[9] S.A.Velinsky,K.R.Kirschke,Design considerations for automated pavement crack sealing machinery[C], Proceedings of the Second International Conference on Applications of Advanced Technologies in Transportation Engineering,18-21 August 1991,pp. 77-80

[10] D. Meignen,M. Bernadet,H. Briand,One Application of Neural Networks for Detection of Defects Using Video Data Bases:Identification of Road Distresses[J],Proceedings Database and Expert Systems Applications, 1997, 9(8):459-464

[11] 李爱霞，管海燕，钟良，等. 基于张量投票的道路表面裂缝检测[J]. 应用科学学报，2015, 33(5):541-549.

[12] Singh, T.R.; Roy, S.; Singh, O.I.; Sinam, T.; Singh, K.M. A New Local Adaptive Thresholding Technique in Binarization. International Journal

of Computer Science Issues 2011, 8.

[13] Bernsen, J.: 'Dynamic thresholding of gray-level images'. Proc. 8th Int. Conf. on Pattern Recognition, Paris, 1986, pp. 1251–1255

[14] Chow, C.K., and Kaneko, T.: 'Automatic detection of the left ventricle from cineangiograms', Comput. Biomed. Res., 1972, 5, pp. 388–410

[15] J. Sauvola and M. Pietikainen, "Adaptive document image binarization," Pattern Recognition 33(2), pp. 225–236, 2000.

[16] Kashani, A.; Olsen, M.; Parrish, C.; Wilson, N. A Review of LIDAR Radiometric Processing: From Ad Hoc Intensity Correction to Rigorous Radiometric Calibration. *SENSORS-BASEL* **2015**, *15*, 28099-28128.

[17] Mordohai, P.; Medioni, G. Tensor Voting: A Perceptual Organization Approach to Computer Vision and Machine Learning. Synthesis Lectures on Image, Video, and Multimedia Processing 2006.

[18] Haiyan Guan, Jonathan Li, Yongtao Yu, Michael Chapman, Hanyun Wang, Cheng Wang, and Ruifang Zhai: Iterative Tensor Voting for Pavement Crack Extraction Using Mobile Laser Scanning Data